\documentclass[11pt]{article}
\usepackage{amssymb,amsmath,amsfonts}
\usepackage{graphicx}
\usepackage{graphics}
\usepackage{eepic,epsfig}

\makeatletter
\@addtoreset{equation}{section}

\makeatother

\begin{document}

\title{Fermionic currents in topologically nontrivial braneworlds}
\author{ S. Bellucci$^{1}$\thanks{%
E-mail: bellucci@lnf.infn.it }, A. A. Saharian$^{2}$\thanks{%
E-mail: saharian@ysu.am }, D. H. Simonyan$^{2}$\thanks{%
E-mail: david.simonyan94@mail.ru }, V. V. Vardanyan$^{2,3,4}$\thanks{%
E-mail: vardanyan@lorentz.leidenuniv.nl } \vspace{0.3cm} \\
\textit{$^1$ INFN, Laboratori Nazionali di Frascati,}\\
\textit{Via Enrico Fermi 40, 00044 Frascati, Italy} \vspace{0.3cm}\\
\textit{$^2$ Department of Physics, Yerevan State University,}\\
\textit{1 Alex Manoogian Street, 0025 Yerevan, Armenia }\vspace{0.3cm}\\
\textit{$^3$ Lorentz Institute for Theoretical Physics, Leiden University, }%
\\
\textit{2333 CA Leiden, The Netherlands }\vspace{0.3cm}\\
\textit{$^4$ Leiden Observatory, Leiden University, } \\
\textit{\ 2300 RA Leiden, The Netherlands}}
\maketitle

\begin{abstract}
We investigate the influence of a brane on the vacuum expectation value
(VEV) of the current density for a charged fermionic field in background of
locally AdS spacetime with an arbitrary number of toroidally compact
dimensions and in the presence of a constant gauge field. Along compact
dimensions the field operator obeys quasiperiodicity conditions with
arbitrary phases and on the brane it is constrained by the bag boundary
condition. The brane is parallel to the AdS boundary and it divides the
space into two regions with different properties for the fermionic vacuum.
In both these regions, the VEVs for the charge density and the components of
the current density along uncompact dimensions vanish. The components along
compact dimensions are decomposed into the brane-free and brane-induced
contributions. The behavior of the latter in various asymptotic regions of
the parameters is investigated. It particular, it is shown that the
brane-induced contribution is mainly located near the brane and vanishes on
the AdS boundary and on the horizon. An important feature is the finiteness
of the current density on the brane. Applications are given to $Z_{2}$%
-symmetric braneworlds of the Randall-Sundrum type with compact dimensions
for two classes of boundary conditions on the fermionic field. For the
second one we show that the contribution of the brane to the current does
not vanish when the location of the brane tends to the AdS boundary. In odd
spacetime dimensions, the fermionic fields realizing two inequivalent
irreducible representations of the Clifford algebra and having equal phases
in the periodicity conditions give the same contribution to the vacuum
current density. Combining the contributions from these fields, the current
density in odd-dimensional $C$-,$P$- and $T$-symmetric models is obtained.
In the special case of three-dimensional spacetime, the corresponding
results are applied for the investigation of the edge effects on the ground
state current density induced in curved graphene tubes by an enclosed
magnetic flux.
\end{abstract}


\bigskip

\section{Introduction}

\label{sec:introd}

In a variety of quantum field-theoretical problems the fields are defined on
a manifold with a boundary and one must take care imposing suitable boundary
conditions on the corresponding hypersurfaces. The boundaries may have
different physical origins. Examples are interfaces between two media with
different electromagnetic properties in condensed matter physics (e.g.,
media with different dielectric permittivities), various sorts of horizons
in gravitational physics and in non-inertial reference frames, boundaries
separating the spatial regions with different gravitational backgrounds (for
example, de Sitter bubbles in Minkowski spacetime), domain walls in the
theory of phase transitions, and branes in higher-dimensional cosmologies
and in string theories. In a number of physical problems the model is
formulated in non-globally hyperbolic manifolds possessing a timelike
boundary at spatial infinity. In order to preserve the information to be
lost to, or gained from, spatial infinity, appropriate boundary conditions
should be imposed. A well known example of this kind is anti-de Sitter (AdS)
spacetime \cite{Avis78}. Another class of conditions imposed on fields
appear in models with compact spatial dimensions. The latter are an inherent
feature of high-energy theories unifying physical interactions, like
Kaluza-Klein and string theories. Depending on the periodicity conditions
along compact dimensions different topologically inequivalent field
configurations may arise \cite{Isha78}. The quantum effects arising from the
nontrivial topological structure of the background spacetime include
symmetry breaking, topological quantum phase transitions, instabilities in
interacting field theories, and topological mass generation. The topological
issues also play an important role in effective theories describing a number
of condensed matter systems \cite{Qi11}.

In the present paper we consider the combined effects of background
gravitational field and of two sorts of boundary conditions on the local
properties of the vacuum state for a charged fermionic field. As the bulk
geometry we take a locally AdS spacetime with an arbitrary number of
toroidally compactified spatial dimensions (in Poincar\'{e} coordinates).
The first kind of boundary condition is related to the presence of a brane
parallel to the AdS boundary and the second one is related to the
compactification of a part of spatial dimensions to a torus. We impose bag
boundary condition on the brane and quasiperiodicity conditions with general
phases along compact dimensions. The results are generalized for a boundary
condition arising in $Z_{2}$-symmetric braneworld models of the
Randall-Sundrum type with extra compact dimensions.

Our choice of AdS spacetime as a local bulk geometry has several
motivations. First of all, AdS spacetime is maximally symmetric and a large
number of problems in quantum field theory on curved backgrounds is exactly
solvable. That is the case in the problem at hand. The corresponding
investigations may help developing the research tools and insights to deal
with less symmetric geometries. The AdS spacetime naturally appears as a
ground state in extended supergravity and Kaluza-Klein theories and also as
the near horizon geometry of the extremal black holes and domain walls.
Moreover, the AdS spacetime has a constant negative curvature and the
related length scale can serve as a regularization parameter for infrared
divergences in interacting quantum field theories without reducing the
number of symmetries \cite{Call90}. The AdS geometry plays a crucial role in
two exciting developments of the last two decades: the gauge/gravity duality
and the braneworld scenario with large extra dimensions (for reviews see
\cite{Ahar00,Maar10}). Braneworlds naturally appear in the string/M-theory
context and provide an interesting alternative to address various problems
in cosmological and particle physics. A number of particularly important
implications of AdS geometry recently appeared in condensed matter physics
(see, e.g., \cite{Pire14}).

Both types of constraints, induced by the presence of boundaries and by the
compactification of spatial dimensions, give rise to the modification of the
spectrum for vacuum fluctuations of quantum fields. As a result, the vacuum
expectation values (VEVs) of physical quantities are shifted by an amount
depending on the bulk and boundary geometries, and also on the boundary
conditions imposed. This is the familiar Casimir effect (for reviews see
\cite{Most97}). The vacuum energy and the forces acting on the boundaries
were among the main physical quantities of interest in the studies of this
effect. In particular, motivated by the radion stabilization in braneworld
models of the Randall-Sundrum type, the investigations of these quantities
in the geometry of two parallel branes in AdS spacetime have attracted a
great deal of attention (see, for instance, the references in \cite%
{Teo10,Eliz13}). In particular, the fermionic Casimir effect has been
considered in \cite{Eliz13,Flac01,Flac01b,Teo13} (for a recent discussion of
the renormalised fermion expectation values on AdS spacetime in the absence
of branes see, for example, \cite{Ambr15}). The vacuum energy, the Casimir
forces and the VEV of the energy-momentum tensor in higher-dimensional
generalizations of the AdS spacetime with compact internal spaces have been
investigated in \cite{Flac03}.

As another important local characteristic of the vacuum state for charged
fields, bilinear in the field operator, here we consider the VEV of the
current density for a fermionic field in background of locally AdS spacetime
with compact dimensions in the presence of a brane. For a flat background
geometry with an arbitrary number of toroidally compact dimensions, the zero
and finite temperature expectation values of the charge and current
densities for scalar and fermionic fields were investigated in Refs. \cite%
{Beze13c,Bell10,Bell14}. The corresponding results for a special case of a
three-dimensional spacetime with one and two compact spatial dimensions have
been applied to the electronic subsystem of cylindrical and toroidal carbon
nanotubes described within the framework of the effective Dirac model. The
influence of additional boundaries on the vacuum charges and currents with
applications to finite length carbon nanotubes is studied in \cite%
{Bell13,Bell15}. This is the analog of the Casimir effect for the charge and
current densities. The VEVs of the current densities for scalar and Dirac
spinor fields in de Sitter and AdS spacetimes with toroidally compact
subspace have been discussed in \cite{Bell13b} and \cite{Beze15,Bell17},
respectively, for scalar and fermionic fields. The effects of the branes in
background of locally AdS bulk on the VEV of the current density for a
scalar field with general curvature coupling parameter are investigated in
\cite{Bell15b,Bell16}. The general case of the Robin boundary conditions on
the branes was discussed and applications were given to Randall-Sundrum type
braneworlds.

The organization of the present paper is as follows. In the next section we
describe the bulk geometry and the fields under consideration. The boundary
and periodicity conditions are specified for a fermionic field. A complete
set of positive and negative energy fermionic modes are described in Section %
\ref{sec:Modes}. By using the corresponding mode functions, in Section \ref%
{sec:Lreg} we investigate the brane-induced effects on the current density
along compact dimensions in the region between the brane and AdS boundary
(L-region). The behavior of the current density in different asymptotic
regions of the parameters is discussed in detail. Similar investigations for
the region between the brane and AdS horizon (R-region) are presented in
Section \ref{sec:Rreg}. In Section \ref{sec:Brane}, the vacuum currents are
considered in $Z_{2}$-symmetric braneworlds with a single brane and with an
arbitrary number of toroidally compact spatial dimensions for two types of
boundary conditions on the brane. The numerical results are given for the
simplest generalization of the Randall-Sundrum model with a single extra
compact dimension. The fermionic current density in parity and time-reversal
symmetric models in odd-dimensional spacetimes is considered in Section \ref%
{sec:OddD}. Applications are given to deformed carbon nanotubes described
within the framework of the effective Dirac model in three-dimensional
spacetime. The main results of the paper are summarized in Section \ref%
{sec:Conc}.

\section{Background geometry and the fields}

\label{sec:Geom}

The background geometry we consider is described by the $(D+1)$-dimensional
line element
\begin{equation}
ds^{2}=e^{-2y/a}\eta _{ik}dx^{i}dx^{k}-dy^{2},  \label{ds2b}
\end{equation}%
where $i,k=0,1,\ldots ,D-1$ and $\eta _{ik}=\mathrm{diag}(1,-1,\ldots ,-1)$
is the Minkowskian metric tensor in $D$-dimensional subspace with the
coordinates $(x^{0}=t,x^{1},\ldots ,x^{D-1})$. The local geometrical
characteristics corresponding to (\ref{ds2b}) coincide with those for AdS
spacetime with the curvature radius $a$. In particular, for the curvature
scalar and the Ricci tensor one has $R=-D(D+1)/a^{2}$ and $R_{\mu \nu
}=-Dg_{\mu \nu }/a^{2}$. However, the global geometry we shall be concerned
about is different. Namely, it will be assumed that the spatial dimension $%
x^{i}$, $i=p+1,\ldots ,D-1$, is compactified to a circle with the length $%
L_{i}$, $0\leqslant x^{i}\leqslant L_{i}$. For the remaining coordinates one
has $-\infty <x^{i}<+\infty $, $i=1,\ldots ,p$, and $-\infty <y<+\infty $.
Hence, in the problem at hand the subspace $(x^{1},\ldots ,x^{D-1})$ has the
topology $R^{p}\times T^{q}$, $q=D-p-1$, where $T^{q}$ stands for a $q$%
-dimensional torus (for a discussion of physical effects in models with
toroidal dimensions, see \cite{Eliz95}). For the further consideration, it
is convenient, in addition to the coordinate $y$, to use the conformal
coordinate $z$, defined as $z=ae^{y/a}$ with the range $0\leqslant z<\infty $%
. In terms of the latter, the metric tensor is written in a conformally flat
form
\begin{equation}
g_{\mu \nu }=(a/z)^{2}\mathrm{diag}\left( 1,-1,\ldots ,-1\right) ,
\label{gmunu}
\end{equation}%
with the spacetime coordinates $x^{\mu }=(x^{0},x^{1},\ldots
,x^{D-1},x^{D}=z)$. The hypersurfaces $z=0$ and $z=\infty $ correspond to
the AdS boundary and horizon, respectively. Note that for the proper length
of the $i$th compact dimension, measured by an observer with a fixed
coordinate $z$, one has $L_{(p)i}=(a/z)L_{i}$.

We are interested in combined effects of the nontrivial topology and
boundaries on the local characteristics of the vacuum state $\left\vert
0\right\rangle $ for a massive fermionic field $\psi (x)$. Assuming the
presence of an external abelian gauge field $A_{\mu }(x)$, the corresponding
field equation reads
\begin{equation}
i\gamma ^{\mu }\left( \partial _{\mu }+\Gamma _{\mu }+ieA_{\mu }\right) \psi
(x)-m\psi (x)=0,  \label{FieldEq}
\end{equation}%
where $e$ is the coupling between the fermionic and gauge fields and $\Gamma
_{\mu }$ is the spin connection. For the curved spacetime Dirac matrices one
has $\gamma ^{\mu }=e_{(b)}^{\mu }\gamma ^{(b)}$, with $\gamma ^{(b)}$ being
the corresponding flat spacetime matrices and $e_{(b)}^{\mu }$ are the
tetrad fields. For a fermionic field realizing the irreducible
representation of the Clifford algebra the matrices $\gamma ^{(b)}$ are $%
N\times N$ matrices with $N=2^{[(D+1)/2]}$, where the square brackets stand
for the integer part of the enclosed expression. For odd $D$ the irreducible
representation is unique up to a similarity transformation, whereas for even
$D$ there are two inequivalent irreducible representations (see Section \ref%
{sec:OddD} below). In the conformal coordinates $x^{\mu }$, with the metric
tensor (\ref{gmunu}), we can take the tetrad fields in the form $%
e_{(b)}^{\mu }=(z/a)\delta _{b}^{\mu }$. The corresponding spin connection
has the components $\Gamma _{k}=\eta _{kl}\gamma ^{(D)}\gamma ^{(l)}/(2z)$
for $k=0,\ldots ,D-1$, and $\Gamma _{D}=0$.

In the discussion below we assume the presence of a boundary, parallel to
the AdS boundary and located at $z=z_{0}$, on which the field operator is
constrained by the bag boundary condition
\begin{equation}
(1+i\gamma ^{\mu }n_{\mu })\psi (x)=0,\;z=z_{0},  \label{Bagbc}
\end{equation}%
where $n_{\mu }$ is the corresponding normal. The respective value of the $y$%
-coordinate we shall denote by $y_{0}$, $y_{0}=a\ln (z_{0}/a)$. Note that
the physical distance from the boundary is given by $|y-y_{0}|$. Though the
boundary under consideration my have different physical origins (for
example, in carbon nanotubes it corresponds to the edge of the tube), for
the convenience of the discussion below we shall use the term 'brane'. It
divides the background space into two regions: $0\leqslant z\leqslant z_{0}$
and $z\geqslant z_{0}$. We shall refer to them as L- and R-regions (left and
right regions), respectively. For the normal one has $n_{\mu }=\delta _{\mu
}^{D}a/z$ in the L-region and $n_{\mu }=-\delta _{\mu }^{D}a/z$ in the
R-region. From (\ref{Bagbc}) it follows that the normal component of the
fermionic current vanishes on the brane. This feature is used in bag models
of hadrons for confinement of quarks. Note that, though the geometrical
characteristics of the background geometry do not depend on $z$, the
boundary under consideration has a nonzero extrinsic curvature tensor with
nonzero components $K_{ik}=\pm g_{ik}/a$, where the upper and lower signs
correspond to the L- and R-regions, respectively. Related to this, the
physical properties of the vacuum will be different in these regions.

The topology of the background space is nontrivial and, in addition to the
boundary condition at $z=z_{0}$, we need to specify the periodicity
conditions imposed on the field operator along compact dimensions. For the
spatial dimension $x^{l}$, $l=p+1,\ldots ,D-1$, we take the quasiperiodicity
condition
\begin{equation}
\psi (t,x^{1},\ldots ,x^{l}+L_{l},\ldots ,x^{D})=e^{i\alpha _{l}}\psi
(t,x^{1},\ldots ,x^{l},\ldots ,x^{D}),  \label{PerCondb}
\end{equation}%
with a constant phase $\alpha _{l}$. The special cases of the most
frequently used conditions with $\alpha _{l}=0$ and $\alpha _{l}=\pi $
correspond to untwisted and twisted fields. As for the gauge field, we
assume the simplest configuration with $A_{\mu }=\mathrm{const}$. The
corresponding effects on quantum properties of the vacuum are of the
Aharonov-Bohm type and they are related to the nontrivial topology of the
background space. The components of the vector potential along noncompact
dimensions are simply removed by a gauge transformation and only the
components along compact dimensions are physically relevant. Hence, our
model is specified by the set of parameters $\{\alpha _{l},A_{l}\}$ with $%
l=p+1,\ldots ,D-1$.

Under the gauge transformation of the field variables $\psi (x)=\psi
^{\prime }(x)e^{-ie\chi }$, $A_{\mu }=A_{\mu }^{\prime }+\partial _{\mu
}\chi $, with the function $\chi =b_{\mu }x^{\mu }$, we obtain a new set of
parameters $\{\alpha _{l}^{\prime },A_{l}^{\prime }\}=\{\alpha
_{l}+eb_{l}L_{l},A_{l}-b_{l}\}$. In particular, in the gauge with $b_{\mu
}=A_{\mu }$ the vector potential vanishes and for the new phases in the
quasiperiodicity conditions for the field $\psi ^{\prime }(x)$ one gets%
\begin{equation}
\tilde{\alpha}_{l}=\alpha _{l}+eA_{l}L_{l}.  \label{alftilde}
\end{equation}%
Hence, the effects of $\alpha _{l}$ and $A_{l}$ are not physically
independent: the physical effects depend on these parameters in the form of
the combination (\ref{alftilde}) which is invariant under the gauge
transformation. In particular, a constant gauge field induces nontrivial
effective phases for twisted and untwisted fields and vice versa: the
nontrivial phases can be interpreted in terms of a constant gauge filed (or
in terms of the magnetic flux). In what follows we will work in the gauge $%
(\psi ^{\prime }(x),A_{\mu }^{\prime }=0)$ omitting the primes. Along the $l$%
th compact dimension the field $\psi ^{\prime }(x)$ obeys the condition (\ref%
{PerCondb}) with $\alpha _{l}$ replaced by $\tilde{\alpha}_{l}$ from (\ref%
{alftilde}). The part in the definition of the latter coming from the vector
potential can be interpreted in terms of the magnetic flux $\Phi _{l}$
enclosed by the $l$th dimension: $eA_{l}L_{l}=-2\pi \Phi _{l}/\Phi _{0}$
(the minus sign comes from the fact that $A_{l}$ is the covariant component
of the $(D+1)$-vector and it is related to the $l$th component of the
spatial vector $\mathbf{A}$ by $A_{l}=-\mathbf{A}_{l}$), with $\Phi
_{0}=2\pi /e$ being the flux quantum. Of course, this flux is fictive, it
lives in the embedding space. However, it can be real flux if the model
under consideration is realized as a brane in a higher dimensional
spacetime. Another problem where the magnetic flux $\Phi _{l}$ has the real
physical sense will be considered in Section \ref{sec:OddD}.

\section{Fermionic modes}

\label{sec:Modes}

The VEVs of physical observables bilinear in the field operator are
expressed in terms of the sums over a complete set of positive and negative
energy fermionic modes $\{\psi _{\beta }^{(+)},\psi _{\beta }^{(-)}\}$,
where the set of quantum numbers $\beta $ specifies the solution. These
modes obey the field equation (\ref{FieldEq}) (with $A_{\mu }=0$ in the
gauge under consideration), the boundary condition (\ref{Bagbc}) and the
quasiperiodicity conditions (\ref{PerCondb}) with $\alpha _{l}$ replaced by $%
\tilde{\alpha _{l}}$. In order to find the solutions to the field equation
one needs to specify the representation of the flat spacetime Dirac matrices
(for the construction of the Dirac matrices in an arbitrary number of
spacetime dimensions see, for example, \cite{Park09}). We find it convenient
to use the representation (see also \cite{Bell17})
\begin{eqnarray}
\gamma ^{(0)} &=&\left(
\begin{array}{cc}
0 & \chi _{0} \\
\chi _{0}^{\dagger } & 0%
\end{array}%
\right) ,\;\gamma ^{(D)}=si\,\left(
\begin{array}{cc}
1 & 0 \\
0 & -1%
\end{array}%
\right) ,\;s=\pm 1,  \notag \\
\gamma ^{(l)} &=&\left(
\begin{array}{cc}
0 & \chi _{l} \\
-\chi _{l}^{\dagger } & 0%
\end{array}%
\right) ,\;l=1,2,\ldots ,D-1,  \label{gaml}
\end{eqnarray}%
with $N/2\times N/2$ matrices $\chi _{0}$, $\chi _{l}$. In even dimensional
spacetimes the irreducible representation is unique (up to a similarity
transformation) and one can take $s=1$. In odd-dimensional spacetimes, the
values $s=+1$ and $s=-1$ correspond to two inequivalent irreducible
representations of the Clifford algebra. From the anticommutation relations
for the Dirac matrices we obtain the following relations%
\begin{eqnarray}
\chi _{l}\chi _{n}^{\dagger }+\chi _{n}\chi _{l}^{\dagger } &=&2\delta
_{nl},\;\chi _{l}^{\dagger }\chi _{n}+\chi _{n}^{\dagger }\chi _{l}=2\delta
_{nl},  \notag \\
\chi _{0}\chi _{l}^{\dagger } &=&\chi _{l}\chi _{0}^{\dagger },\;\chi
_{0}^{\dagger }\chi _{l}=\chi _{l}^{\dagger }\chi _{0},\;\chi _{0}^{\dagger
}\chi _{0}=1,  \label{Relxi}
\end{eqnarray}%
with $l,n=1,2,\ldots ,D-1$. In the special case $D=2$, taking $\chi
_{0}=\chi _{1}=1$, we get $\gamma ^{(0)}=\sigma _{\mathrm{P}1}$, $\gamma
^{(1)}=i\sigma _{\mathrm{P}2}$, $\gamma ^{(2)}=si\sigma _{\mathrm{P}3}$,
where $\sigma _{\mathrm{P}\mu }$ are the Pauli matrices.

With the flat spacetime matrices (\ref{gaml}), substituting in the field
equation (\ref{FieldEq}) the Dirac matrices $\gamma ^{\mu }=(z/a)\delta
_{b}^{\mu }\gamma ^{(b)}$, the complete set of the positive and negative
energy solutions of the field equation can be found in a way similar to that
we have described in Appendix of Ref. \cite{Bell17}. In accordance with the
symmetry of the problem, the dependence of the mode functions on the
coordinates $(t,\mathbf{x})=(t,x^{1},\ldots ,x^{D-1})$ can be taken in the
form of plane waves $e^{i\mathbf{kx}-i\omega t}$, $\mathbf{kx}%
=\sum_{l=1}^{D-1}k_{i}x^{i}$, with the momentum $\mathbf{k}=(k_{1},\ldots
,k_{D-1})$ and the energy $\omega $. The mode functions are presented as%
\begin{eqnarray}
\psi _{\beta }^{(+)}(x) &=&z^{\frac{D+1}{2}}e^{i\mathbf{kx}-i\omega t}\left(
\begin{array}{c}
\frac{\mathbf{k\chi }\chi _{0}^{\dagger }+i\lambda -\omega }{\omega }%
Z_{ma+s/2}(\lambda z)w^{(\sigma )} \\
i\chi _{0}^{\dagger }\frac{\mathbf{k\chi }\chi _{0}^{\dagger }+i\lambda
+\omega }{\omega }Z_{ma-s/2}(\lambda z)w^{(\sigma )}%
\end{array}%
\right) ,  \notag \\
\psi _{\beta }^{(-)}(x) &=&z^{\frac{D+1}{2}}e^{i\mathbf{kx}+i\omega t}\left(
\begin{array}{c}
i\chi _{0}\frac{\mathbf{k\chi }^{\dagger }\chi _{0}-i\lambda +\omega }{%
\omega }Z_{ma+s/2}(\lambda z)w^{(\sigma )} \\
\frac{\mathbf{k\chi }^{\dagger }\chi _{0}-i\lambda -\omega }{\omega }%
Z_{ma-s/2}(\lambda z)w^{(\sigma )}%
\end{array}%
\right) ,  \label{Modesb}
\end{eqnarray}%
where $w^{(\sigma )}$, $\sigma =$ $1,\ldots ,N/2$, are one-column matrices
having $N/2$ rows and the elements $w_{l}^{(\sigma )}=\delta _{l\sigma }$.
In (\ref{Modesb}), $\mathbf{k\chi }=\sum_{l=1}^{D-1}k_{l}\chi _{l}$, $%
0\leqslant \lambda <\infty $, $\omega =\sqrt{\lambda ^{2}+k^{2}}$, $k=|%
\mathbf{k}|$ and
\begin{equation}
Z_{\nu }(u)=c_{1}J_{\nu }(u)+c_{2}Y_{\nu }(u),  \label{Znu}
\end{equation}%
is a linear combination of the Bessel and Neumann functions $J_{\nu }(u)$
and $Y_{\nu }(u)$. The coefficients $c_{1}$ and $c_{2}$ depend on the region
under consideration and will be determined below separately in the L- and
R-regions.

For the components of the momentum along uncompact spatial dimensions, as
usual, one has $-\infty <k_{i}<+\infty $, $i=1,\ldots ,p$. The eigenvalues
of the components along compact dimensions are quantized by the periodicity
conditions:%
\begin{equation}
k_{l}=\frac{2\pi n_{l}+\tilde{\alpha _{l}}}{L_{l}},\;n_{l}=0,\pm 1,\pm
2,\ldots ,  \label{kl}
\end{equation}%
where $l=p+1,\ldots ,D-1$. For $\tilde{\alpha _{l}}=2\pi p_{l}$, with $p_{l}$
being an integer, the parameter $\tilde{\alpha _{l}}$ is removed from the
problem by the redefinition of the quantum number $n_{l}$. Therefore, only
the fractional part of $\tilde{\alpha _{l}}/2\pi $ is physically relevant.
The set of quantum numbers $\beta $ specifying the modes is given by $\beta
=(\lambda ,\mathbf{k}_{(p)},\mathbf{n}_{q},\sigma )$, where $\mathbf{k}%
_{(p)}=(k_{1},\ldots ,k_{p})$ is the momentum in the non-compact subspace
and $\mathbf{n}_{q}=(n_{p+1},\ldots ,n_{D-1})$ determines the momentum in
the compact subspace. The orthonormalization condition for the mode
functions reads
\begin{equation}
\int d^{D}x\,(a/z)^{D}\psi _{\beta }^{(\pm )\dagger }\psi _{\beta ^{\prime
}}^{(\pm )}=\delta _{\beta \beta ^{\prime }},  \label{Norm}
\end{equation}%
where $\delta _{\beta \beta ^{\prime }}$ is understood as the Dirac delta
function for the continuous components of $\beta $ and the Kronecker delta
for discrete ones.

We are interested in the VEV of the current density $j^{\mu }=e\bar{\psi}%
\gamma ^{\mu }\psi $, where for the Dirac conjugate one has $\bar{\psi}=\psi
^{\dagger }\gamma ^{(0)}$. Expanding the field operator in terms of the
complete set of modes and using the anticommutation relations for the
annihilation and creation operators, the VEV $\left\langle 0\right\vert
j^{\mu }\left\vert 0\right\rangle \equiv \langle j^{\mu }\rangle $ is
presented in the form of the mode sum%
\begin{equation}
\langle j^{\mu }(x)\rangle =\frac{e}{2}\sum_{\beta }\left[ \bar{\psi}_{\beta
}^{(-)}(x)\gamma ^{\mu }\psi _{\beta }^{(-)}(x)-\bar{\psi}_{\beta
}^{(+)}(x)\gamma ^{\mu }\psi _{\beta }^{(+)}(x)\right] .  \label{jVEV}
\end{equation}%
Here, $\sum_{\beta }$ stands for the integration over the continuous
components of the collective index $\beta $ and for the summation over the
discrete components. The functions $Z_{\nu }(u)$ in (\ref{Modesb}) and the
eigenvalues for $\lambda $ are different in the L- and R-regions and we
investigate the corresponding current densities separately.

\section{Current density in the L-region}

\label{sec:Lreg}

First we consider the region between the brane and the AdS boundary,
corresponding to $0\leqslant z\leqslant z_{0}$. In the range of the mass $%
ma\geq 1/2$ and for $c_{2}\neq 0$ in (\ref{Znu}) the modes (\ref{Modesb})
are not normalizable. Hence, for this range, from the normalizability
condition it follows that $c_{2}=0$ and the mode functions are given by (\ref%
{Modesb}) with $Z_{\nu }(u)=c_{1}J_{\nu }(u)$. From the boundary condition (%
\ref{Bagbc}) it follows that the eigenvalues for the quantum number $\lambda
$ are roots of the equation
\begin{equation*}
J_{ma-1/2}(\lambda z_{0})=0,
\end{equation*}%
for both the cases $s=\pm 1$. We shall denote the corresponding positive
roots with respect to $\lambda z_{0}$ by $\lambda _{n}=\lambda
_{n}(ma)=\lambda z_{0}$, $n=1,2,\ldots $, assuming that they are numerated
in the ascending order, $\lambda _{n+1}>\lambda _{n}$. Note that the roots $%
\lambda _{n}$ do not depend on the location of the brane.

Now the mode functions are written as%
\begin{eqnarray}
\psi _{\beta }^{(+)}(x) &=&C_{L\beta }^{(+)}z^{\frac{D+1}{2}}e^{i\mathbf{kx}%
-i\omega t}\left(
\begin{array}{c}
\frac{\mathbf{k\chi }\chi _{0}^{\dagger }+i\lambda -\omega }{\omega }%
J_{ma+s/2}(\lambda z)w^{(\sigma )} \\
i\chi _{0}^{\dagger }\frac{\mathbf{k\chi }\chi _{0}^{\dagger }+i\lambda
+\omega }{\omega }J_{ma-s/2}(\lambda z)w^{(\sigma )}%
\end{array}%
\right) ,  \notag \\
\psi _{\beta }^{(-)}(x) &=&C_{L\beta }^{(-)}z^{\frac{D+1}{2}}e^{i\mathbf{kx}%
+i\omega t}\left(
\begin{array}{c}
i\chi _{0}\frac{\mathbf{k\chi }^{\dagger }\chi _{0}-i\lambda +\omega }{%
\omega }J_{ma+s/2}(\lambda z)w^{(\sigma )} \\
\frac{\mathbf{k\chi }^{\dagger }\chi _{0}-i\lambda -\omega }{\omega }%
J_{ma-s/2}(\lambda z)w^{(\sigma )}%
\end{array}%
\right) ,  \label{ModesbL}
\end{eqnarray}%
where $\lambda =\lambda _{n}/z_{0}$. For a massless field one has $\lambda
_{n}=\pi (n-1/2)$.

The normalization coefficients $C_{L\beta }^{(\pm )}$ are determined from
the condition (\ref{Norm}), where the integration over $z$ is done in the
region $[0,z_{0}]$ and on the right-hand side the Kronecker delta $\delta
_{\lambda _{n}\lambda _{n^{\prime }}}=\delta _{nn^{\prime }}$ appears. By
using the standard integral for the square of the Bessel function one finds%
\begin{equation}
|C_{\beta }^{(\pm )}|^{2}=\frac{J_{ma+1/2}^{-2}(\lambda _{n})}{2(2\pi
)^{p}V_{q}a^{D}z_{0}^{2}},  \label{Cbpm}
\end{equation}%
where $V_{q}=L_{p+1}\cdots L_{D-1}$ is the volume of the compact subspace.
As seen, the normalization constants are the same for both the
representations $s=1$ and $s=-1$.

For the range of masses $0\leq ma<1/2$ the modes with $c_{2}\neq 0$ in (\ref%
{Znu}) are normalizable. In this case, in order to determine the additional
coefficient in the mode function one needs to specify a boundary condition
for the field on the AdS boundary. This kind of boundary conditions for
fermions have been discussed, for example, in Refs. \cite%
{Brei82b,Amse09,Ahar11}. Here we shall consider a special type of boundary
condition when the bag boundary condition is imposed on the hypersurface $%
z=z_{1}>0$ and then the limiting transition $z_{1}\rightarrow 0$ is taken.
As it will be shown in the next section, this procedure leads to the mode
functions which are given by (\ref{ModesbL}) for all $ma\geq 0$.

We start our investigation for $\langle j^{\mu }\rangle $ with the charge
density corresponding to the component $\mu =0$. Plugging the modes (\ref%
{ModesbL}) in (\ref{jVEV}) we get
\begin{equation}
\langle j^{0}\rangle =\frac{2e}{a}z^{D+2}\sum_{\beta }|C_{\beta }^{(\pm
)}|^{2}\frac{1}{\omega }w^{(\sigma )\dagger }\mathbf{k\chi }^{\dagger }\chi
_{0}w^{(\sigma )}\left[ J_{ma+s/2}^{2}(\lambda z)-J_{ma-s/2}^{2}(\lambda z)%
\right] ,  \label{j0}
\end{equation}%
where%
\begin{equation}
\sum_{\beta }=\sum_{\mathbf{n}_{q}}\int d\mathbf{k}_{(p)}\sum_{n=1}^{\infty
}\sum_{\sigma =1}^{N/2}\,,  \label{Sumbet}
\end{equation}%
with $\mathbf{k}_{(p)}=(k_{1},\ldots ,k_{p})$ being the momentum in the
uncompact subspace. Now we note that for a $N/2\times N/2$ matrix $M$ one
has $\sum_{\sigma =1}^{N/2}w^{(\sigma )\dagger }Mw^{(\sigma )}=\mathrm{tr}M$%
. By taking into account the relations (\ref{Relxi}) it can be seen that $%
\mathrm{tr}(\chi _{l}^{+}\chi _{0})=0$ and, hence, $\sum_{\sigma
=1}^{N/2}w^{(\sigma )\dagger }\mathbf{k\chi }^{\dagger }\chi _{0}w^{(\sigma
)}=0$. From here we conclude that the VEV\ of the charge density vanishes.

Now we turn to the $l$th spatial component of the current density. By using
the mode sum (\ref{jVEV}) with the modes (\ref{ModesbL}), in a way similar
to that for the charge density we can see that%
\begin{equation}
\langle j^{l}\rangle =-\frac{(2\pi )^{-p}Nez^{D+2}}{2V_{q}a^{D+1}z_{0}}\sum_{%
\mathbf{n}_{q}}\int d\mathbf{k}_{(p)}k_{l}\sum_{n=1}^{\infty }\frac{%
\sum_{j=\pm 1}J_{ma+js/2}^{2}(\lambda _{n}z/z_{0})}{\sqrt{\lambda
_{n}^{2}+z_{0}^{2}k^{2}}J_{ma+1/2}^{2}(\lambda _{n})}.  \label{jlb}
\end{equation}%
For $l=1,\ldots ,p$, the integrand is an odd function with respect to the
momentum $k_{l}$ and the corresponding component of the current density is
zero, $\langle j^{l}\rangle =0$. Hence, a nonzero current density may appear
along the compact dimensions only. This is a purely topological effect of
the Aharonov-Bohm type and is induced by the nontrivial phases in the
quasiperiodicity conditions (or, alternatively, by the enclosed magnetic
fluxes). For $\tilde{\alpha _{l}}=2\pi p_{l}$, with $p_{l}$ being an
integer, after passing to the summation over $n_{l}^{\prime }=n_{l}+p_{l}$,
we see that the contributions in (\ref{jlb}) coming from the modes with
positive and negative values of $k_{l}$ cancel each other and the resulting
current density vanishes. Another important conclusion following from (\ref%
{jlb}) is that the current densities for the representations with $s=1$ and $%
s=-1$ coincide. We will continue the investigation of the current density in
the L-region for the case $s=1$.

The representation (\ref{jlb}) contains the eigenvalues $\lambda _{n}$ which
are given implicitly, as the zeros of the Bessel function. In order to
obtain a representation more convenient for the asymptotic and numerical
analysis, and for explicit extraction of the brane-induced contribution, we
apply to the series over $n$ a variant of the generalized Abel-Plana formula
\cite{Saha87}%
\begin{eqnarray}
\sum_{n=1}^{\infty }\frac{f(\lambda _{n})}{\lambda
_{n}J_{ma+1/2}^{2}(\lambda _{n})} &=&\frac{1}{2}\int_{0}^{\infty }du\,f(u)-%
\frac{1}{2\pi }\int_{0}^{\infty }du\,\frac{K_{ma-1/2}(u)}{I_{ma-1/2}(u)}
\notag \\
&&\times \left[ e^{\left( 1/2-ma\right) \pi i}f(iu)+e^{\left( ma-1/2\right)
\pi i}f(-iu)\right] ,  \label{SummAP}
\end{eqnarray}%
valid for a function $f(u)$ analytic in the right half-plane of the complex
variable $u$ (function $f(u)$ may have branch points on the imaginary axis,
for the conditions imposed on this function see \cite{Saha87}). In (\ref%
{SummAP}), $I_{\nu }(u)$ and $K_{\nu }(u)$ are the modified Bessel
functions. In the problem under consideration the function $f(u)$ is
specified as%
\begin{equation}
f(u)=\frac{u}{\sqrt{u^{2}+z_{0}^{2}k^{2}}}\left[
J_{ma+1/2}^{2}(uz/z_{0})+J_{ma-1/2}^{2}(uz/z_{0})\right] ,  \label{fu}
\end{equation}%
and has branch points $u=\pm iz_{0}k$.

After application of formula (\ref{SummAP}) to the series over $n$ in (\ref%
{jlb}) and integration over the angular coordinates of the vector $\mathbf{k}%
_{(p)}$, the VEV\ of the current density is decomposed as%
\begin{equation}
\langle j^{l}\rangle =\langle j^{l}\rangle _{0}+\langle j^{l}\rangle _{b},
\label{jdec}
\end{equation}%
where the term
\begin{equation}
\langle j^{l}\rangle _{0}=-\frac{(4\pi )^{-p/2}Nez^{D+2}}{2\Gamma
(p/2)V_{q}a^{D+1}}\sum_{\mathbf{n}_{q}}\int_{0}^{\infty
}dk_{(p)}\,k_{(p)}^{p-1}k_{l}\int_{0}^{\infty }d\lambda \,\lambda \frac{%
J_{ma+s/2}^{2}(\lambda z)+J_{ma-s/2}^{2}(\lambda z)}{\sqrt{\lambda ^{2}+k^{2}%
}},  \label{jl0}
\end{equation}%
comes from the first integral in the right-hand side of (\ref{SummAP}) and
coincides with the current density in the geometry without the brane (see
\cite{Bell17}). The term
\begin{eqnarray}
\langle j^{l}\rangle _{b} &=&-\frac{4\left( 4\pi \right) ^{-p/2-1}Nez^{D+2}}{%
\Gamma (p/2)V_{q}a^{D+1}}\sum_{\mathbf{n}_{q}}k_{l}\int_{0}^{\infty
}dk_{(p)}\,k_{(p)}^{p-1}\int_{k}^{\infty }du\,  \notag \\
&&\times \frac{u}{\sqrt{u^{2}-k^{2}}}\frac{K_{ma-1/2}(uz_{0})}{%
I_{ma-1/2}(uz_{0})}\left[ I_{ma+1/2}^{2}(uz)-I_{ma-1/2}^{2}(uz)\right] ,
\label{jlL}
\end{eqnarray}%
is the contribution induced by the brane. For a fixed $z$, the latter goes
to zero in the limit $z_{0}\rightarrow \infty $.

The current density $\langle j^{l}\rangle _{0}$ in the brane-free geometry
has been investigated in \cite{Bell17}. An alternative representation is
given by
\begin{equation}
\langle j^{l}\rangle =-\frac{eNa^{-D-1}L_{l}}{(2\pi )^{(D+1)/2}}%
\sum_{n_{l}=1}^{\infty }n_{l}\sin (\tilde{\alpha}_{l}n_{l})\sum_{\mathbf{n}%
_{q-1}}\,\cos (\tilde{\mathbf{\alpha }}_{q-1}\cdot \mathbf{n}%
_{q-1})\sum_{j=0,1}q_{ma-j}^{\frac{D+1}{2}}\left( 1+\sum_{i=p+1}^{D-1}\frac{%
n_{i}^{2}L_{i}^{2}}{2z^{2}}\right) ,  \label{jl02}
\end{equation}%
where, $\mathbf{n}_{q-1}=(n_{p+1},\ldots ,n_{l-1},n_{l+1},\ldots ,n_{D-1})$,
$\tilde{\mathbf{\alpha }}_{q-1}\cdot \mathbf{n}_{q-1}=\sum_{i=1,\neq l}^{D-1}%
\tilde{\alpha}_{i}n_{i}$. The function $q_{\nu }^{\mu }(x)$ is expressed in
terms of the hypergeometric function $F(a,b;c;x)$ as%
\begin{equation}
q_{\nu }^{\mu }(x)=\frac{\sqrt{\pi }\Gamma (\nu +\mu +1)}{2^{\nu +1}\Gamma
(\nu +3/2)x^{\nu +\mu +1}}F\left( \frac{\nu +\mu +1}{2},\frac{\nu +\mu +2}{2}%
;\nu +\frac{3}{2};\frac{1}{x^{2}}\right) .  \label{qmunu}
\end{equation}%
Note that for $\mu =1,2,\ldots $ (this corresponds to odd values of $D$ in (%
\ref{jl02})) one has $q_{\nu }^{\mu }(x)=(-1)^{\mu }\partial _{x}^{\mu
}Q_{\nu }(x)$ with $Q_{\nu }(x)$ being the Legendre function of the second
kind. For $\mu =1/2,3/2,\ldots $ (even values of $D$ in (\ref{jl02})) the
function $q_{\nu }^{\mu }(x)$ is expressed in terms of the elementary
functions. In what follows we will be mainly concerned about the
brane-induced effects in the current density.

By taking into account that $k^{2}=k_{(p)}^{2}+k_{(q)}^{2}$, with
\begin{equation}
k_{(q)}^{2}=\sum_{i=p+1}^{D-1}\left( 2\pi n_{i}+\tilde{\alpha}_{i}\right)
^{2}/L_{i}^{2},  \label{kq2}
\end{equation}%
the contribution (\ref{jlL}) is further simplified by using the relation%
\begin{equation}
\int_{0}^{\infty }dk_{(p)}\,k_{(p)}^{p-1}\int_{k}^{\infty }du\frac{ug(u)}{%
\sqrt{u^{2}-k^{2}}}=\frac{\sqrt{\pi }\Gamma \left( p/2\right) }{2\Gamma
\left( (p+1)/2\right) }\int_{k_{(q)}}^{\infty }du\,u\left(
u^{2}-k_{(q)}^{2}\right) ^{\frac{p-1}{2}}g(u),  \label{relint}
\end{equation}%
for a given function $g(u)$. This leads to the following expression for the
brane-induced contribution to the current density:
\begin{eqnarray}
\langle j^{l}\rangle _{b} &=&-\frac{NeA_{p}z^{D+2}}{V_{q}a^{D+1}}\sum_{%
\mathbf{n}_{q}}k_{l}\int_{k_{(q)}}^{\infty }du\,u\left(
u^{2}-k_{(q)}^{2}\right) ^{\frac{p-1}{2}}  \notag \\
&&\times \frac{K_{ma-1/2}(uz_{0})}{I_{ma-1/2}(uz_{0})}\left[
I_{ma+1/2}^{2}(uz)-I_{ma-1/2}^{2}(uz)\right] ,  \label{jlLb}
\end{eqnarray}%
with the notation%
\begin{equation}
A_{p}=\frac{\left( 4\pi \right) ^{-(p+1)/2}}{\Gamma \left( (p+1)/2\right) }.
\label{Cp}
\end{equation}%
Note that the integrand in (\ref{jlLb}) is always negative. Both the
brane-free and brane-induced contributions in the $l$th component of the
vacuum current density are odd periodic functions of the phase $\tilde{\alpha%
}_{l}$ and even periodic functions of $\tilde{\alpha}_{i}$ with $i\neq l$,
with the period $2\pi $. In particular, they are periodic functions of the
magnetic flux with the period equal to the flux quantum $\Phi _{0}$. The
charge flux through the hypersurface $x^{l}=\mathrm{const}$ is given by $%
n_{l}^{(l)}\langle j^{l}\rangle $, where $n_{i}^{(l)}=\delta _{i}^{l}a/z$ is
the normal to that hypersurface. The product $a^{D}n_{l}^{(l)}\langle
j^{l}\rangle $ depends on the variables having the dimension of length in
the form of the dimensionless combinations $z_{0}/z$, $L_{i}/z$, $ma$. This
feature is a consequence of the maximal symmetry of the AdS spacetime. Note
that the ratio $L_{i}/z=L_{(p)i}/a$ is the proper length of the $i$th
compact dimension in units of the curvature radius $a$.

In order to further clarify the behavior of the current density we pass to
the investigation of the VEV (\ref{jlLb}) in special cases and in various
asymptotic regions of the parameters. First we consider the current density
of a massless fermionic field. In this case the modified Bessel functions in
(\ref{jlLb}) are expressed in terms of the elementary functions and one gets%
\begin{eqnarray}
\langle j^{l}\rangle _{b} &=&\frac{2NeA_{p}z^{D+1}}{V_{q}a^{D+1}}\sum_{%
\mathbf{n}_{q}}k_{l}\int_{k_{(q)}}^{\infty }du\,\frac{\left(
u^{2}-k_{(q)}^{2}\right) ^{\frac{p-1}{2}}}{e^{2uz_{0}}+1}  \notag \\
&=&-\frac{2^{-3p/2}eNz^{D+1}}{\pi ^{p/2+1}V_{q}a^{D+1}z_{0}^{p}}%
\sum_{n=1}^{\infty }\frac{(-1)^{n}}{n^{p}}\sum_{\mathbf{n}%
_{q}}k_{l}g_{p/2}(2nz_{0}k_{(q)}),  \label{jlbm0}
\end{eqnarray}%
with the function%
\begin{equation}
g_{\nu }(x)=x^{\nu }K_{\nu }(x).  \label{gnu}
\end{equation}%
The second representation in (\ref{jlbm0}) is obtained from the first one by
using the expansion $1/(e^{x}+1)=-\sum_{n=1}^{\infty }(-1)^{n}e^{-nx}$. The
massless fermionic field is conformally invariant in an arbitrary number of
spatial dimensions and the result (\ref{jlbm0}) is obtained from the
expression for the current density in the region between two boundaries at $%
z=0$ and $z=z_{0}$ on a locally Minkowskian bulk with compact dimensions $%
(x^{p+1},\ldots ,x^{D-1})$ by using the conformal relation $\langle
j^{l}\rangle _{b}=(z/a)^{D+1}\langle j^{l}\rangle _{b}^{(M)}$ . Note that
the boundary $z=0$ in the Minkowski bulk is the conformal image of the AdS
boundary. We can see that $\langle j^{l}\rangle _{b}^{(M)}$ obtained from (%
\ref{jlbm0}) coincides with the result from \cite{Bell10} (the sign
difference is related to the fact that $\tilde{\alpha}_{i}$ in \cite{Bell10}
corresponds to $-\tilde{\alpha}_{i}$ in the present paper).

The Minkowskian limit corresponds to $a\rightarrow \infty $ for fixed $y$
and $y_{0}$. In this limit the conformal coordinates $z$ and $z_{0}$ are
large, $z\approx a+y$, $z-z_{0}\approx y-y_{0}$, and, consequently, both the
order and the argument of the modified Bessel functions in (\ref{jlLb}) are
large. By using the corresponding uniform asymptotic expansions \cite{Abra72}%
, to the leading order we get
\begin{equation}
\langle j^{l}\rangle _{b}^{(M)}=\frac{NeA_{p}m}{V_{q}}\sum_{\mathbf{n}%
_{q}}k_{l}\int_{m_{(q)}}^{\infty }dx\,\left( x^{2}-m_{(q)}^{2}\right) ^{%
\frac{p-1}{2}}\frac{e^{-2x|y-y_{0}|}}{x+m},  \label{jlbM}
\end{equation}%
with the notation $m_{(q)}=\sqrt{m^{2}+k_{(q)}^{2}}$. This expression
coincides with the result from \cite{Bell13} for a boundary in a flat bulk
with topology $R^{p+1}\times T^{q}$ (again, with the sign difference related
to definition of the parameters $\tilde{\alpha}_{i}$). For a massless field
the current density induced by a single boundary in flat spacetime vanishes.

Now let us consider the behavior of the current density near the AdS
boundary and near the brane for the fixed location of the brane. For points
close to the AdS boundary one has $z/z_{0}\ll 1$ and the main contribution
to the integral in (\ref{jlLb}) comes from the region of the integration
where the argument of the functions $I_{ma\pm 1/2}(uz)$ is small. By using
the corresponding asymptotic expression, the leading order term reads%
\begin{equation}
\langle j^{l}\rangle _{b}\approx \frac{NeA_{p}a^{-D-1}z^{D+2ma+1}}{%
2^{2ma-1}V_{q}\Gamma ^{2}(ma+1/2)}\sum_{\mathbf{n}_{q}}k_{l}\int_{k_{(q)}}^{%
\infty }du\,u^{2ma}\left( u^{2}-k_{(q)}^{2}\right) ^{\frac{p-1}{2}}\frac{%
K_{ma-1/2}(uz_{0})}{I_{ma-1/2}(uz_{0})},  \label{NearbL}
\end{equation}%
and on the AdS boundary the brane-induced contribution vanishes as $%
z^{D+2ma+1}$. Note that the brane-free contribution behaves in a similar
manner, $\langle j^{l}\rangle _{0}\propto z^{D+2ma+1}$.

The representation (\ref{jlLb}) for the current density is not well suited
for the investigation of the near-brane asymptotic. In order to obtain an
alternative representation, we apply to the series over $n_{l}$ in the
initial expression (\ref{jlb}) the Abel--Plana-type formula \cite{Bell10}%
\begin{eqnarray}
&&\frac{2\pi }{L_{l}}\sum_{n_{l}=-\infty }^{\infty
}g(k_{l})f(|k_{l}|)=\int_{0}^{\infty }du[g(u)+g(-u)]f(u)  \notag \\
&&\qquad +i\int_{0}^{\infty }du\,[f(iu)-f(-iu)]\sum_{j=\pm 1}\frac{g(iju)}{%
e^{uL_{l}+ij\tilde{\alpha}_{l}}-1},  \label{AP}
\end{eqnarray}%
for given functions $g(u)$, $f(u)$ and with $k_{l}$ defined in (\ref{kl})
(formula (\ref{AP}) is reduced to the standard Abel-Plana formula in the
special case $g(x)=1$, $\tilde{\alpha}_{l}=0$). For the series in (\ref{jlb}%
) one has $g(u)=u$ and the first integral in (\ref{AP}) is zero. By making
use of the relation%
\begin{equation}
\sum_{j=\pm 1}\frac{j}{e^{uL_{l}+ij\tilde{\alpha}_{l}}-1}=\frac{2}{i}%
\sum_{r=1}^{\infty }e^{-ruL_{l}}\sin (r\tilde{\alpha}_{l}),  \label{sumj}
\end{equation}%
in the last term in (\ref{AP}), the integral over $u$ is expressed in terms
of the modified Bessel function $K_{1}(nL_{l}\sqrt{\lambda
^{2}+k^{2}-k_{l}^{2}})$. Evaluating the remaining integral over $\mathbf{k}%
_{(p)}$ by using the formula from \cite{Prud86}, the VEV\ of the current
density is presented as (as it has been shown above, the current densities
for the representations $s=1$ and $s=-1$ are the same and we consider the
case $s=1$)%
\begin{eqnarray}
\langle j^{l}\rangle  &=&-\frac{2Nea^{-D-1}z^{D+2}}{(2\pi
)^{p/2+1}V_{q}L_{l}^{p}z_{0}^{2}}\sum_{r=1}^{\infty }\frac{\sin (r\tilde{%
\alpha}_{l})}{r^{p+1}}\sum_{\mathbf{n}_{q-1}}\sum_{n=1}^{\infty }  \notag \\
&&\times g_{p/2+1}(rL_{l}\sqrt{\lambda _{n}^{2}/z_{0}^{2}+k_{(q-1)}^{2}})%
\frac{\sum_{j=\pm 1}J_{ma+j/2}^{2}(\lambda _{n}z/z_{0})}{J_{ma+1/2}^{2}(%
\lambda _{n})},  \label{jlAlt}
\end{eqnarray}%
where the function $g_{\nu }(x)$ is defined by (\ref{gnu}) and
\begin{equation}
k_{(q-1)}^{2}=\sum_{i=p+1,\neq l}^{D-1}\left( 2\pi n_{i}+\tilde{\alpha}%
_{i}\right) ^{2}/L_{i}^{2}.  \label{kq-1}
\end{equation}%
Note that in the representation (\ref{jlAlt}) the terms of the series over $n
$ decay exponentially for large $\lambda _{n}$. In the case of a massless
field we have $\lambda _{n}=\pi (n-1/2)$ and the ratio of the Bessel
functions in (\ref{jlAlt}) is equal to $z_{0}/z$. In this case we get the
standard conformal relation with the corresponding representation of the
current density between two boundaries in locally Minkowskian spacetime with
compact dimensions.

The total current, per unit surface along the uncompact dimensions, is
obtained by integration of (\ref{jlAlt}):%
\begin{eqnarray}
V_{q}\int_{0}^{z_{0}}dz\,\sqrt{|g|}\langle j^{l}\rangle &=&-\frac{2Ne}{(2\pi
)^{p/2+1}L_{l}^{p}}\sum_{r=1}^{\infty }\frac{\sin (r\tilde{\alpha}_{l})}{%
r^{p+1}}  \notag \\
&&\times \sum_{\mathbf{n}_{q-1}}\sum_{n=1}^{\infty }g_{p/2+1}(rL_{l}\sqrt{%
\lambda _{n}^{2}/z_{0}^{2}+k_{(q-1)}^{2}}).  \label{jlInt}
\end{eqnarray}%
Note that the dependence on the curvature radius of the background
spacetime, on the mass of the field and on the location of the brane appears
through the ratio $\lambda _{n}/z_{0}$. We recall that the roots $\lambda
_{n}$ are completely determined by the parameter $ma$ and do not depend on
the location of the brane.

In the model with a single compact dimension $x^{l}$ with the length $L_{l}$
($q=1$, $l=D-1$) the formula (\ref{jlAlt}) is specified to%
\begin{equation}
\langle j^{l}\rangle |_{q=1}=-\frac{2Nea^{-D-1}z^{D+2}}{\left( 2\pi \right)
^{D/2}L_{l}^{D-1}z_{0}^{2}}\sum_{r=1}^{\infty }\frac{\sin (r\tilde{\alpha}%
_{l})}{r^{D-1}}\sum_{n=1}^{\infty }\frac{g_{D/2}\left( r\lambda
_{n}L_{l}/z_{0}\right) }{J_{ma+1/2}^{2}(\lambda _{n})}\sum_{j=\pm
1}J_{ma+j/2}^{2}(\lambda _{n}z/z_{0}).  \label{jlAltq1}
\end{equation}%
An alternative expression in this special case is obtained from (\ref{jlLb}):%
\begin{eqnarray}
\langle j^{l}\rangle _{b}|_{q=1} &=&-Ne\frac{A_{D-2}z^{D+2}}{a^{D+1}L_{l}}%
\sum_{n_{l}=-\infty }^{+\infty }k_{l}\int_{|k_{l}|}^{\infty }du\,u\left(
u^{2}-k_{l}^{2}\right) ^{\frac{D-3}{2}}  \notag \\
&&\times \frac{K_{ma-1/2}(uz_{0})}{I_{ma-1/2}(uz_{0})}\left[
I_{ma+1/2}^{2}(uz)-I_{ma-1/2}^{2}(uz)\right] .  \label{jlq1}
\end{eqnarray}%
In this and in the next sections, for numerical investigations of the
current density we consider the special case $D=4$ with a single compact
dimension of the length $L_{l}=L$ and with the phase in the periodicity
condition $\tilde{\alpha}_{l}=\tilde{\alpha}$. For this model the
corresponding formulas are obtained from (\ref{jlAltq1}) and (\ref{jlq1})
taking $p=2$ and $q=1$.

\begin{figure}[tbph]
\begin{center}
\epsfig{figure=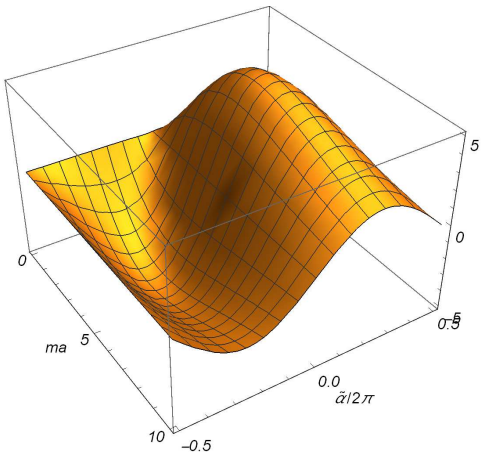,width=7.cm,height=6.cm}
\end{center}
\caption{The brane-induced charge flux along the compact dimension, $%
a^{D}n_{l}^{(l)}\langle j^{l}\rangle _{b}$, versus the mass and the phase in
the periodicity condition for a $D=4$ model with a single compact dimension
of the length $L$. The graph is plotted for $z/z_{0}=0.95$ and $L/z_{0}=1/3$%
. }
\label{fig1}
\end{figure}

In figure \ref{fig1} we have plotted the dependence of the quantity $%
a^{D}n_{l}\langle j^{l}\rangle _{b}/e$, with $n_{l}=a/z$ (for the example at
hand $l=3$), on the phase $\tilde{\alpha}$ and on the mass of the field (in
units of the inverse curvature radius) for fixed values $z/z_{0}=0.95$ and $%
L/z_{0}=1/3$. Recall that $n_{l}\langle j^{l}\rangle _{b}$ is the charge
flux through the spatial hypersurface $x^{l}=\mathrm{conts}$, induced by the
brane. The current density is a periodic function of $\tilde{\alpha}/2\pi $
with the period 1 and we have plotted the dependence for the interval $%
-0.5\leq \tilde{\alpha}/2\pi \leq 0.5$. Unlike the case of the R-region (see
below), the brane-induced current density in the L-region does not vanish
for a massless filed. With an initial increase of the mass the absolute
value of the current density increases and after passing its maximum value
tends to zero for large masses. In the latter limit the orders of the
modified Bessel functions in (\ref{jlLb}) are large and we can use the
corresponding uniform asymptotic expansions. The contribution of the modes
with $k_{(q)}\gg m$ and $k_{(q)}(z_{0}-z)\gg 1$ is suppressed by the factor $%
\exp [-2k_{(q)}\left( z_{0}-z\right) ]$. The dominant contribution comes
from the modes with $k_{(q)}(z_{0}-z)\lesssim 1$ and the brane-induced VEV
for large masses decays as $(z/z_{0})^{2ma}$, or as $\exp [-2ma(y_{0}-y)]$
in terms of the proper distance from the brane.

An important conclusion made from the representation (\ref{jlAlt}) is that
the current density is finite on the brane and the corresponding value can
be directly obtained putting $z=z_{0}$:%
\begin{equation}
\langle j^{l}\rangle _{z=z_{0}}=-\frac{2Nea^{-D-1}z_{0}^{D}}{(2\pi
)^{p/2+1}V_{q}L_{l}^{p}}\sum_{r=1}^{\infty }\frac{\sin (r\tilde{\alpha}_{l})%
}{r^{p+1}}\sum_{\mathbf{n}_{q-1}}\sum_{n=1}^{\infty }g_{p/2+1}(rL_{l}\sqrt{%
\lambda _{n}^{2}/z_{0}^{2}+k_{(q-1)}^{2}}).  \label{jlz0}
\end{equation}%
In this sense, the behavior of the current density is essentially different
from that for the VEVs of the fermion condensate and the energy-momentum
tensor. The latter contain surface divergences well known from quantum field
theory on manifolds with boundaries (see, for instance, \cite{Most97}). The
absence of the surface divergences in the VEV of the current density in the
problem at hand can be understood as follows (see also a similar feature for
scalar currents in \cite{Bell15b,Bell16}). The divergences in the VEVs of
local observables are completely determined by the local geometrical
characteristics of the bulk and boundary geometries. The compactification we
have considered does not change these characteristics and they are the same
as those for the problem in AdS spacetime without compact dimension. But in
the latter problem the current density vanishes and, consequently contains
no divergences. The toroidal compactification leaves the local geometry
unchanged and, hence, no additional divergences will arise as a consequence
of that. Comparing (\ref{jlz0}) with (\ref{jlInt}), we obtain the following
relation between the current density on the brane and the current integrated
over the coordinate $z$:%
\begin{equation}
\int_{0}^{z_{0}}dz\,\sqrt{|g|}\langle j^{l}\rangle =\frac{a^{D+1}}{z_{0}^{D}}%
\langle j^{l}\rangle _{z=z_{0}}.  \label{reltot}
\end{equation}

Introducing in (\ref{jlLb}) a new integration variable $x=uz_{0}$ we see
that the resulting integral is a function of two combinations $z_{0}k_{(q)}$
and $z/z_{0}$. For $zk_{(q)}\gg 1$, the arguments of the modified Bessel
functions in the integrand are large. By using the corresponding asymptotics
we find%
\begin{equation}
\langle j^{l}\rangle _{b}\approx \frac{NeA_{p}mz^{D}}{V_{q}a^{D}}\sum_{%
\mathbf{n}_{q}}k_{l}\int_{k_{(q)}}^{\infty }du\,\frac{\left(
u^{2}-k_{(q)}^{2}\right) ^{\frac{p-1}{2}}}{ue^{2u\left( z_{0}-z\right) }}.
\label{jlblim1}
\end{equation}%
If, in addition, one has the condition $\left( z_{0}-z\right) k_{(q)}\gg 1$,
the contribution of the corresponding modes to the brane-induced current
density is suppressed by the factor $e^{-2k_{(q)}\left( z_{0}-z\right) }$.
For $z_{0}-z\gg L_{i}$ that condition is obeyed for all the modes and the
integral in (\ref{jlblim1}) is dominated by the contribution from the region
near the lower limit and by the mode with the smallest value for $k_{(q)}$.
Assuming that $|\tilde{\alpha _{i}}|<\pi $, this mode corresponds to $%
n_{i}=0 $ for $i=p+1,\ldots ,D-1$, and the leading order term is presented
as
\begin{equation}
\langle j^{l}\rangle _{b}\approx \frac{Nem\tilde{\alpha}%
_{l}z^{D}(k_{(q)}^{(0)})^{(p-3)/2}e^{-2(z_{0}-z)k_{(q)}^{(0)}}}{2\left( 4\pi
\right) ^{(p+1)/2}V_{q}L_{l}a^{D}\left( z_{0}-z\right) ^{(p+1)/2}},
\label{jlblim2}
\end{equation}%
where
\begin{equation}
k_{(q)}^{(0)2}=\sum_{i=p+1}^{D-1}\tilde{\alpha}_{i}^{2}/L_{i}^{2}.
\label{kq0}
\end{equation}%
From here we conclude that the brane-induced contribution is mainly
localized near the brane in the region $z_{0}-z\lesssim L_{i}$. Figure \ref%
{fig2} displays the brane-induced current density as a function of $z/z_{0}$
for different values of the ratio $z_{0}/L$ (figures near the curves). The
graphs are plotted for the same model as in figure \ref{fig1} and for fixed
values of $ma=2$ and $\tilde{\alpha}=\pi /2$. Note that in terms of the
proper distance from the brane one has $z/z_{0}=e^{-(y_{0}-y)/a}$. For the
example presented, near the AdS boundary, $z\rightarrow 0$, the VEV decays
as $(z/z_{0})^{9}$.

\begin{figure}[tbph]
\begin{center}
\epsfig{figure=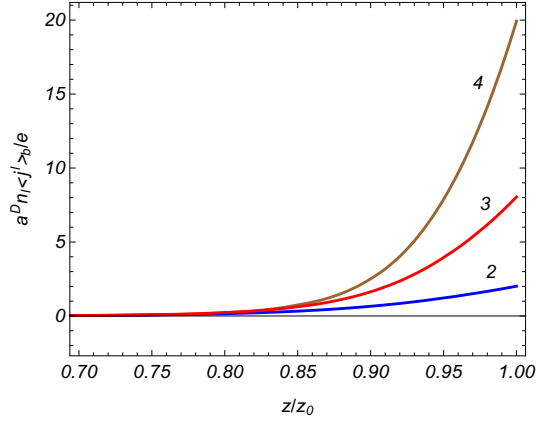,width=7.cm,height=5.5cm}
\end{center}
\caption{The dependence of the brane-induced current density in the L-region
on the ratio $z/z_{0}$. The graphs are plotted for different values of $%
z_{0}/L$ (numbers near the curves) and for fixed $ma=2$ and $\tilde{\protect%
\alpha}=\protect\pi /2$.}
\label{fig2}
\end{figure}

Now let us consider the VEV of the current density as a function of the
location of the brane. When the brane is close to the AdS horizon ($z_{0}$
is large compared to the other length scales of the problem) the integral in
(\ref{jlLb}) is dominated by the region where the argument $uz_{0}$ of the
modified Bessel functions is large. By using the respective asymptotic
formulas, we can see that the contribution of the mode with a given $\mathbf{%
n}_{q}$ is suppressed by the factor $e^{-2k_{(q)}z_{0}}$. From here it
follows that the dominant contribution comes from the mode with the lowest
possible value for $k_{(q)}$. Under the assumption $|\tilde{\alpha _{i}}%
|<\pi $, this mode corresponds to $n_{i}=0$, $i=p+1,\ldots ,D-1$, with $%
k_{(q)}=k_{(q)}^{(0)}$. To the leading order one finds
\begin{equation}
\langle j^{l}\rangle _{b}\approx -\frac{Ne\tilde{\alpha}%
_{l}z^{D+2}k_{(q)}^{(0)(p+1)/2}e^{-2k_{(q)}^{(0)}z_{0}}}{2^{p+2}\pi
^{(p-1)/2}V_{q}L_{l}a^{D+1}z_{0}^{(p+1)/2}}\left[
I_{ma+1/2}^{2}(k_{(q)}^{(0)}z)-I_{ma-1/2}^{2}(k_{(q)}^{(0)}z)\right] .
\label{jlNearHor}
\end{equation}%
Hence, when the brane is close to the AdS horizon, the brane-induced current
density is suppressed by the factor $e^{-2k_{(q)}^{(0)}z_{0}}$.

If the brane is close to the AdS boundary, both $z_{0}$ and $z$ are small.
In the investigation of the corresponding asymptotics it is convenient to
use the representation (\ref{jlAlt}). The dominant contribution to the
series over $\mathbf{n}_{q-1}$ comes from the terms with large values of $%
|n_{i}|$ with $i\neq l$ and we can replace the summation by the integration
in accordance with%
\begin{equation}
\sum_{\mathbf{n}_{q-1}}f(k_{(q-1)})\rightarrow 2A_{q-2}\frac{V_{q}}{L_{l}}%
\int_{0}^{\infty }dx\,x^{q-2}f(x),  \label{Repl1}
\end{equation}%
where $k_{(q-1)}^{2}$ is defined by (\ref{kq-1}). After the integration over
$x$ with the help of the formula
\begin{equation}
\int_{0}^{\infty }dx\,x^{\beta -1}g_{\nu }(c\sqrt{x^{2}+b^{2}})=\frac{%
2^{\beta /2-1}}{c^{\beta }}\Gamma \left( \beta /2\right) g_{\nu +\beta
/2}\left( bc\right) ,  \label{Intf1}
\end{equation}%
one gets $\langle j^{l}\rangle \approx \langle j^{l}\rangle |_{q=1}$, where $%
\langle j^{l}\rangle |_{q=1}$ is given by (\ref{jlAltq1}). In the limit
under consideration the argument of the function $g_{D/2}\left( r\lambda
_{n}L_{l}/z_{0}\right) $ is large and we use the corresponding asymptotic
formula $g_{\nu }(x)\approx \sqrt{\pi /2}x^{\nu -1/2}e^{-x}$. The dominant
contribution comes from the lowest mode $r=1$ and we get%
\begin{equation}
\langle j^{l}\rangle \approx -\frac{Nez^{D+2}\sin (\tilde{\alpha}%
_{l})\lambda _{1}^{(D-1)/2}e^{-\lambda _{1}L_{l}/z_{0}}}{(2\pi
)^{(D-1)/2}a^{D+1}L_{l}^{(D-1)/2}z_{0}^{(D+3)/2}}\frac{\sum_{j=\pm
1}J_{ma+j/2}^{2}(\lambda _{1}z/z_{0})}{J_{ma+1/2}^{2}(\lambda _{1})},
\label{Nearb}
\end{equation}%
with the VEV suppressed by the factor $e^{-\lambda _{1}L_{l}/z_{0}}$.

For small values of $L_{l}$ compared with the lengths of the remaining
compact dimensions, the dominant contribution to (\ref{jlLb}) comes from the
modes with large values of $|n_{i}|$, $i\neq l$. In this case, to the
leading order, we can replace the summation over $\mathbf{n}_{q-1}$ by the
integration in accordance with (\ref{Repl1}). With this replacement, instead
of $u$ we introduce a new integration variable $w$ as $u=\sqrt{%
x^{2}+w^{2}+k_{l}^{2}}$ and then polar coordinates in the plane $(x,w)$.
After the integration over the angular variable, we get%
\begin{equation}
\langle j^{l}\rangle _{b}\approx \langle j^{l}\rangle _{b}|_{q=1},
\label{SmallL}
\end{equation}%
with $\langle j^{l}\rangle _{b}|_{q=1}$ from (\ref{jlq1}). The same relation
takes place for the brane-free parts. If in addition $L_{l}\ll z_{0}$, the
arguments of the modified Bessel functions in the integrand are large and we
employ the corresponding asymptotic formulas. To the leading order this gives%
\begin{equation}
\langle j^{l}\rangle _{b}\approx Nem\frac{A_{D-2}z^{D}}{a^{D}L_{l}}%
\sum_{n_{l}=-\infty }^{+\infty }k_{l}|k_{l}|^{D-3}\int_{1}^{\infty }du\,%
\frac{\left( u^{2}-1\right) ^{(D-3)/2}}{ue^{2u\left( z_{0}-z\right) |k_{l}|}}%
.  \label{SmallL1}
\end{equation}%
For a massless field this leading term vanishes. The expression on the right
of (\ref{SmallL1}) is further simplified under the condition $L_{l}\ll
z_{0}-z$. In this case the dominant contribution comes from the integration
region near the lower limit and from the mode with the minimal value for $%
|k_{l}|$. Assuming $|\tilde{\alpha}_{l}|<\pi $, this mode corresponds to $%
n_{l}=0$ and we get%
\begin{equation}
\langle j^{l}\rangle _{b}\approx \frac{\mathrm{sgn}(\tilde{\alpha}_{l})Nem|%
\tilde{\alpha}_{l}|^{\left( D-3\right) /2}}{2^{D}\pi
^{(D-1)/2}a^{D}L_{l}^{\left( D-1\right) /2}}\frac{z^{D}e^{-2\left(
z_{0}-z\right) |\tilde{\alpha}_{l}|/L_{l}}}{\left( z_{0}-z\right) ^{\left(
D-1\right) /2}}.  \label{SmallL2}
\end{equation}%
In this limit the sign of $\langle j^{l}\rangle _{b}/e$ coincides with that
for $\tilde{\alpha}_{l}$.

\begin{figure}[tbph]
\begin{center}
\begin{tabular}{cc}
\epsfig{figure=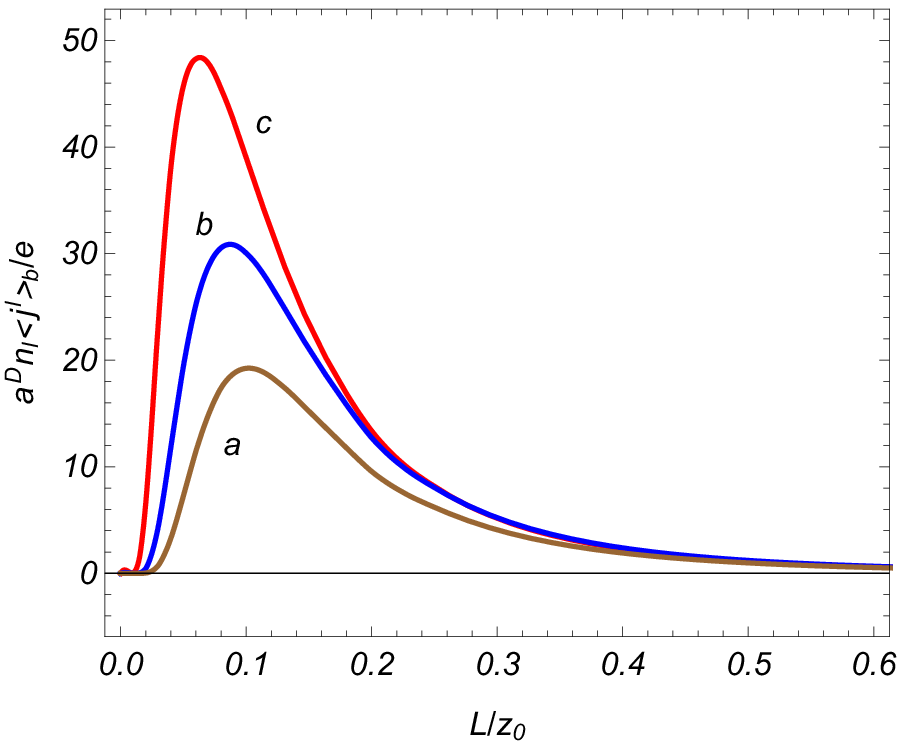,width=7.cm,height=5.5cm} & \quad %
\epsfig{figure=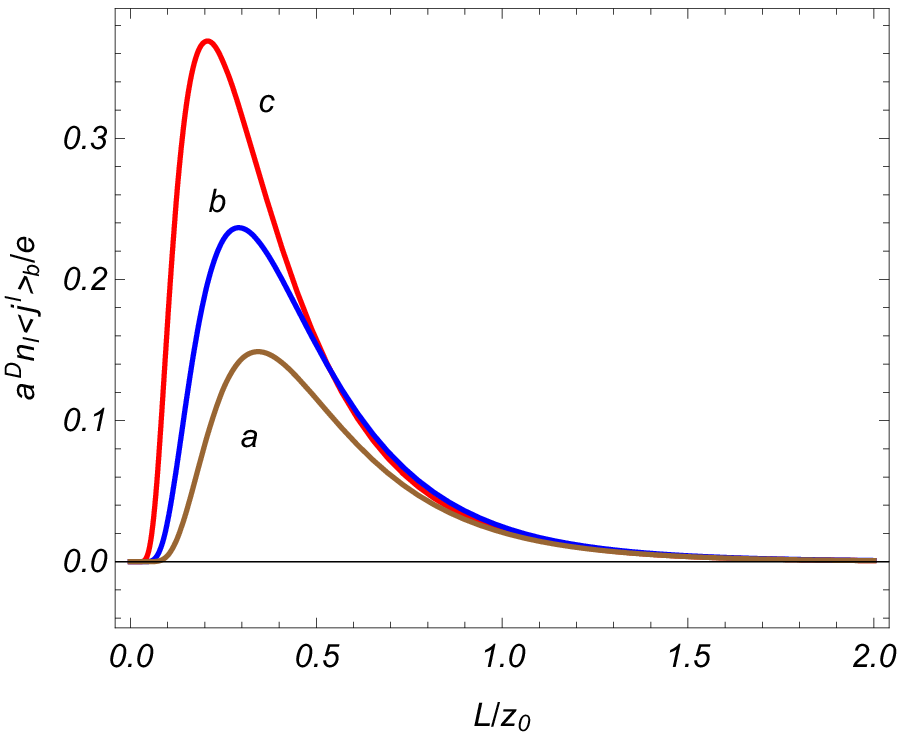,width=7.cm,height=5.5cm}%
\end{tabular}%
\end{center}
\caption{The brane-induced charge flux along the compact dimension, $%
a^{D}n_{l}^{(l)}\langle j^{l}\rangle $, as a function of the rescaled length
$L/z_{0}$ for separate values of the phase $\tilde{\protect\alpha}=2\protect%
\pi /3$ (a), $\tilde{\protect\alpha}=\protect\pi /2$ (b), $\tilde{\protect%
\alpha}=\protect\pi /3$ (c). For the mass we have taken $ma=2$. For the left
and right panels $z/z_{0}=0.95$ and $z/z_{0}=0.8$, respectively.}
\label{fig3}
\end{figure}

In the opposite limit of large values $L_{l}$ it is convenient to use the
representation (\ref{jlAlt}). The argument of the function $g_{p/2+1}(x)$ is
large and we can use the corresponding asymptotic expression. The dominant
contribution to the current density comes from the modes with $n=r=1$ and
with the lowest value of $k_{(q-1)}$. Denoting the latter by $%
k_{(q-1)}^{(0)} $, to the leading order one finds%
\begin{equation}
\langle j^{l}\rangle =-\frac{Nea^{-D-1}z^{D+2}\sin (\tilde{\alpha}%
_{l})x^{(p+1)/2}e^{-x}}{(2\pi
)^{(p+1)/2}V_{q}L_{l}^{p}z_{0}^{2}J_{ma+1/2}^{2}(\lambda _{1})}\left[
J_{ma+1/2}^{2}(\lambda _{1}z/z_{0})+J_{ma-1/2}^{2}(\lambda _{1}z/z_{0})%
\right] .  \label{largeL}
\end{equation}%
with $x=L_{l}\sqrt{\lambda _{1}^{2}/z_{0}^{2}+k_{(q-1)}^{(0)2}}$. Assuming
that $|\tilde{\alpha _{i}}|<\pi $, one has
\begin{equation}
k_{(q-1)}^{(0)2}=\sum_{i=p+1,\neq l}^{D-1}\tilde{\alpha}_{i}^{2}/L_{i}^{2}.
\label{kqm10}
\end{equation}%
The nontrivial phases along compact dimensions $x^{i}$, $i\neq l$, enhance
the suppression for the current density along the $l$th dimension. In figure %
\ref{fig3}, the current density is plotted as a function of $L/z_{0}$ for
fixed values of $z/z_{0}=0.95$ (left panel) and $z/z_{0}=0.8$ (right panel).
The graphs correspond to $ma=2$ and to the values of the phase $\tilde{\alpha%
}=2\pi /3$ (a), $\tilde{\alpha}=\pi /2$ (b), $\tilde{\alpha}=\pi /3$ (c).
The suppression for small values of $L$, described by (\ref{SmallL2}), is
seen in figure \ref{fig3}.

\section{Currents in the R-region}

\label{sec:Rreg}

In this section we consider the current density in the region between the
brane and the AdS horizon, $z_{0}\leq z<\infty $. The corresponding function
$Z_{\nu }(u)$ is given by (\ref{Znu}). From the boundary condition (\ref%
{Bagbc}) it follows that $Z_{ma+1/2}(\lambda z_{0})=0$. For the ratio of the
coefficients in (\ref{Znu}) this gives
\begin{equation}
\frac{c_{2}}{c_{1}}=-\frac{J_{ma+1/2}(\lambda z_{0})}{Y_{ma+1/2}(\lambda
z_{0})},  \label{crat}
\end{equation}%
and the function $Z_{\nu }(u)$ is expressed in terms of the function
\begin{equation}
g_{\mu ,\nu }(x,u)=J_{\mu }(x)Y_{\nu }(u)-J_{\nu }(u)Y_{\mu }(x).
\label{gemu}
\end{equation}%
In the R-region the spectrum for the quantum number $\lambda $ is continuous
and the corresponding mode functions are written as%
\begin{eqnarray}
\psi _{\beta }^{(+)}(x) &=&C_{R\beta }^{(+)}z^{\frac{D+1}{2}}e^{i\mathbf{kx}%
-i\omega t}\left(
\begin{array}{c}
\frac{\mathbf{k\chi }\chi _{0}^{\dagger }+i\lambda -\omega }{\omega }%
g_{ma+1/2,ma+s/2}(\lambda z_{0},\lambda z)w^{(\sigma )} \\
i\chi _{0}^{\dagger }\frac{\mathbf{k\chi }\chi _{0}^{\dagger }+i\lambda
+\omega }{\omega }g_{ma+1/2,ma-s/2}(\lambda z_{0},\lambda z)w^{(\sigma )}%
\end{array}%
\right) ,  \notag \\
\psi _{\beta }^{(-)}(x) &=&C_{R\beta }^{(-)}z^{\frac{D+1}{2}}e^{i\mathbf{kx}%
+i\omega t}\left(
\begin{array}{c}
i\chi _{0}\frac{\mathbf{k\chi }^{\dagger }\chi _{0}-i\lambda +\omega }{%
\omega }g_{ma+1/2,ma+s/2}(\lambda z_{0},\lambda z)w^{(\sigma )} \\
\frac{\mathbf{k\chi }^{\dagger }\chi _{0}-i\lambda -\omega }{\omega }%
g_{ma+1/2,ma-s/2}(\lambda z_{0},\lambda z)w^{(\sigma )}%
\end{array}%
\right) .  \label{ModesbR}
\end{eqnarray}%
The normalization constants are found from (\ref{Norm}) with the integration
over $z$ in the range $[z_{0},\infty )$ and with $\delta \left( \lambda
^{\prime }-\lambda \right) $ in the right-hand side:%
\begin{equation}
\left\vert C_{R\beta }^{(\pm )}\right\vert ^{2}=\lambda \frac{\left[
J_{ma+1/2}^{2}(\lambda z_{0})+Y_{ma+1/2}^{2}(\lambda z_{0})\right] ^{-1}}{%
4\left( 2\pi \right) ^{p}V_{q}a^{D}}.  \label{CR}
\end{equation}%
Given the mode functions (\ref{ModesbR}), the VEV\ of the current density is
evaluated by using (\ref{jVEV}).

Let us consider the limiting transition $z_{0}\rightarrow 0$ for the mode
functions (\ref{ModesbR}). By taking into account that in this limit%
\begin{equation}
C_{R\beta }^{(+)}g_{ma+1/2,ma\pm 1/2}(\lambda z_{0},\lambda z)\sim -\sqrt{%
\frac{\lambda }{4\left( 2\pi \right) ^{p}V_{q}a^{D}}}J_{ma\pm 1/2}(\lambda z)
\label{limtr}
\end{equation}%
(and the same with $C_{R\beta }^{(-)}$) we get the transition to the modes
in boundary-free AdS spacetime that are analog of the modes (\ref{ModesbL})
in the problem with the absence of the brane. Hence, we confirmed the
statement in the previous section related to the boundary condition on the
AdS boundary for the modes in the range of mass $0\leq ma<1/2$.

Similarly to the case of the L-region, we can see that the charge density
and the components of the current density along uncompact dimensions vanish:
$\langle j^{l}\rangle =0$ for $l=0,1,\ldots ,p,D$. For the component along
the $l$th compact dimension we get%
\begin{eqnarray}
\langle j^{l}\rangle  &=&-\frac{\left( 4\pi \right) ^{-p/2}Nez^{D+2}}{%
2\Gamma (p/2)V_{q}a^{D+1}}\sum_{\mathbf{n}_{q}}k_{l}\int_{0}^{\infty
}dk_{(p)}\,k_{(p)}^{p-1}  \notag \\
&&\times \int_{0}^{\infty }d\lambda \frac{\lambda }{\omega }\frac{%
\sum_{j=\pm 1}g_{ma+1/2,ma+js/2}^{2}(\lambda z_{0},\lambda z)}{%
J_{ma+1/2}^{2}(\lambda z_{0})+Y_{ma+1/2}^{2}(\lambda z_{0})},  \label{jR}
\end{eqnarray}%
with $l=p+1,\ldots ,D-1$. From here it follows that the current densities
for the representations $s=\pm 1$ coincide in the R-region. For the
extraction of the brane-induced contribution we use the identity%
\begin{equation}
\frac{g_{\nu ,\mu }^{2}(x,y)}{J_{\nu }^{2}(x)+Y_{\nu }^{2}(x)}=J_{\mu
}^{2}(y)-\frac{1}{2}\sum_{n=1,2}\frac{J_{\nu }(x)}{H_{\nu }^{(n)}(x)}H_{\mu
}^{(n)2}(y),  \label{ident}
\end{equation}%
where $H_{\nu }^{(n)}(x)$, $n=1,2$, are the Hankel functions. The
contribution coming from the first term in the right-hand side of (\ref%
{ident}) gives the current density in the problem without the brane (see (%
\ref{jl0})) and the VEV\ is decomposed as in (\ref{jdec}). In the
brane-induced part, coming from the last term in (\ref{ident}), we rotate
the integration contour over $\lambda $ by the angle $\pi /2$ ($-\pi /2$)
for the term with $n=1$ ($n=2$). Introducing the modified Bessel functions
and using the relation (\ref{relint}), for the brane-induced contribution to
the $l$th component of the current density we find%
\begin{eqnarray}
\langle j^{l}\rangle _{b} &=&\frac{NeA_{p}z^{D+2}}{V_{q}a^{D+1}}\sum_{%
\mathbf{n}_{q}}k_{l}\int_{k_{(q)}}^{\infty }du\,u\left(
u^{2}-k_{(q)}^{2}\right) ^{\frac{p-1}{2}}  \notag \\
&&\times \frac{I_{ma+1/2}(uz_{0})}{K_{ma+1/2}(uz_{0})}\left[
K_{ma+1/2}^{2}(uz)-K_{ma-1/2}^{2}(uz)\right] .  \label{jlRb}
\end{eqnarray}%
This component is an odd periodic function of $\tilde{\alpha}_{l}$ and an
even periodic function of $\tilde{\alpha}_{i}$, $i\neq l$, with the period $%
2\pi $. The integrand in (\ref{jlRb}) is positive for $u>k_{(q)}$.

For a massless field the brane-induced VEV (\ref{jlRb}) vanishes. This
result could be directly obtained from the conformal relation of the problem
under consideration in the case of massless field to the corresponding
problem for a boundary in the flat spacetime bulk with compact dimensions.
It is known that in the latter problem, for a massless fermionic field, the
boundary-induced contribution vanishes (see \cite{Bell13}). The flat
spacetime limit for a massive field can be obtained in a way similar to that
we have demonstrated for the L-region. The corresponding VEV is given by (%
\ref{jlbM}).

\begin{figure}[tbph]
\begin{center}
\epsfig{figure=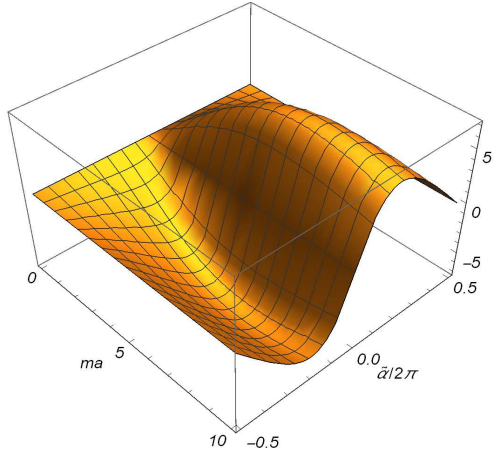,width=7.cm,height=6.cm}
\end{center}
\caption{The same as in figure \protect\ref{fig1} for $z/z_{0}=1.15$
(R-region).}
\label{fig4}
\end{figure}

Figure \ref{fig4} presents the brane-induced current density in the R-region
as a function of the mass and of the phase in the quasiperiodicity condition
for fixed ratios $z_{0}/L=3$ and $z/z_{0}=1.15$. As for figure \ref{fig1},
we consider the model with $D=4$ and $(p,q)=(2,1)$ with the length of the
compact dimension $L$ and the phase $\tilde{\alpha}$. With an initial
increase of the mass the absolute value of the current density increases.
After taking its maximal value the current density exponentially decays for
large values of $ma$. For the modes with $k_{(q)}(z-z_{0})\lesssim 1$ this
decay is like $(z_{0}/z)^{2ma}$.

With a new integration variable $x=uz_{0}$, the resulting integral in (\ref%
{jlRb}) depends on two dimensionless combinations $z_{0}k_{(q)}$ and $%
z/z_{0} $. For $z_{0}k_{(q)}\gg 1$, by using the asymptotics for the
modified Bessel functions for large arguments we can see that the
corresponding leading term is obtained from the one in the L-region, given
by (\ref{jlblim1}), by the replacement $z_{0}-z\rightarrow z-z_{0}$. If
additionally one has the condition $\left( z-z_{0}\right) k_{(q)}\gg 1$, the
contribution of the mode with a given $k_{(q)}$ decays as $%
e^{-2k_{(q)}\left( z-z_{0}\right) }$. In particular, for $z-z_{0}\gg L_{i}$
that condition is valid for all the modes and for the leading term we get
the expression that is obtained from (\ref{jlblim2}) making the replacement $%
z_{0}-z\rightarrow z-z_{0}$. Hence, the leading term is the same for the L-
and R-regions. It is mainly localized near the brane in the region $%
\left\vert z_{0}-z\right\vert \lesssim L_{i}$.

For the investigation of the behavior of the current density in some
asymptotic regions of the parameters we find it convenient to provide
another representation. It is obtained from (\ref{jR}) by using the formula (%
\ref{AP}) for the summation of the series over $n_{l}$. By calculations
similar to those for the L-region one gets%
\begin{eqnarray}
\langle j^{l}\rangle  &=&-\frac{Nea^{-D-1}z^{D+2}}{\left( 2\pi \right)
^{p/2+1}V_{q}L_{l}^{p}}\sum_{r=1}^{\infty }\frac{\sin (r\tilde{\alpha}_{l})}{%
r^{p+1}}\sum_{\mathbf{n}_{q-1}}\int_{0}^{\infty }d\lambda \,\lambda   \notag
\\
&&\times g_{p/2+1}(rL_{l}\sqrt{\lambda ^{2}+k_{(q-1)}^{2}})\frac{\sum_{j=\pm
1}g_{ma+1/2,ma+j/2}^{2}(\lambda z_{0},\lambda z)}{J_{ma+1/2}^{2}(\lambda
z_{0})+Y_{ma+1/2}^{2}(\lambda z_{0})},  \label{jlRalt1}
\end{eqnarray}%
with the function $g_{\nu }(x)$ from (\ref{gnu}). In particular, from this
representation it follows that the current density is finite on the brane
with the value%
\begin{eqnarray}
\langle j^{l}\rangle _{z=z_{0}} &=&-\frac{16Nea^{-D-1}z_{0}^{D}}{\left( 2\pi
\right) ^{p/2+3}V_{q}L_{l}^{p}}\sum_{r=1}^{\infty }\frac{\sin (r\tilde{\alpha%
}_{l})}{r^{p+1}}\sum_{\mathbf{n}_{q-1}}\int_{0}^{\infty }\frac{du}{u}  \notag
\\
&&\times \frac{g_{p/2+1}(rL_{l}\sqrt{u^{2}/z_{0}^{2}+k_{(q-1)}^{2}})}{%
J_{ma+1/2}^{2}(u)+Y_{ma+1/2}^{2}(u)}.  \label{jlz0R}
\end{eqnarray}%
For a massless field, by using the expressions for the functions $J_{\pm
1/2}(x)$ and $Y_{\pm 1/2}(x)$, from (\ref{jlRalt1}) it can be seen that
\begin{equation}
\langle j^{l}\rangle |_{m=0}=-\frac{2Ne(z/a)^{D+1}}{\left( 2\pi \right)
^{(p+3)/2}V_{q}L_{l}^{p+1}}\sum_{r=1}^{\infty }\frac{\sin (r\tilde{\alpha}%
_{l})}{r^{p+2}}\sum_{\mathbf{n}_{q-1}}g_{(p+3)/2}\left(
rL_{l}k_{(q-1)}\right) .  \label{jlRm0}
\end{equation}%
This result is conformally related to the current density in flat spacetime
with toroidally compact dimensions in the absence of boundaries, obtained in
\cite{Bell10} (with the sign difference related to different definitions of $%
\tilde{\alpha}_{i}$; note that in \cite{Bell10} the number of uncompact
dimensions is $p$ instead of $p+1$ as in the present paper). This again
shows that for a massless field the brane-induced contribution in the
R-region vanishes.

Another representation for the current density in the R-region is obtained
from (\ref{jlRalt1}) by using the identity (\ref{ident}) and rotating the
integration contours for the terms with $n=1,2$ in a way similar to that for
(\ref{jlRb}). This leads to the formula%
\begin{eqnarray}
\langle j^{l}\rangle  &=&\langle j^{l}\rangle _{0}+\frac{Nea^{-D-1}z^{D+2}}{%
\left( 2\pi \right) ^{p/2+1}V_{q}L_{l}^{p}}\sum_{r=1}^{\infty }\frac{\sin (r%
\tilde{\alpha}_{l})}{r^{p+1}}\sum_{\mathbf{n}_{q-1}}\int_{k_{(q-1)}}^{\infty
}d\lambda \lambda w_{p/2+1}(rL_{l}\sqrt{\lambda ^{2}-k_{(q-1)}^{2}})  \notag
\\
&&\times \frac{I_{ma+1/2}(\lambda z_{0})}{K_{ma+1/2}(\lambda z_{0})}\left[
K_{ma+1/2}^{2}(\lambda z)-K_{ma-1/2}^{2}(\lambda z)\right] ,  \label{jlRalt2}
\end{eqnarray}%
with the function
\begin{equation}
w_{\nu }(x)=x^{\nu }J_{\nu }(x).  \label{w}
\end{equation}%
The second term in the right-hand side of (\ref{jlRalt2}) is the
contribution induced by the brane ($\langle j^{l}\rangle _{b}$ in the
notation used before).

\begin{figure}[tbph]
\begin{center}
\epsfig{figure=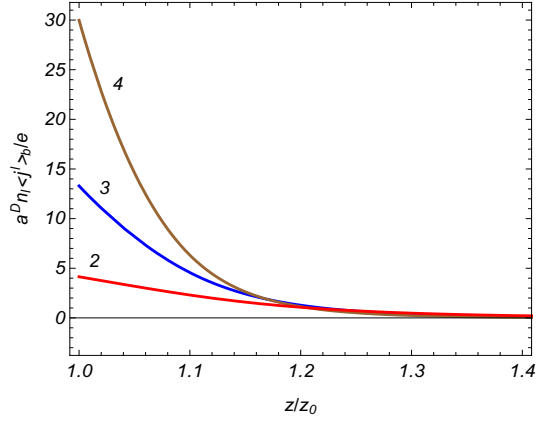,width=7.cm,height=5.5cm}
\end{center}
\caption{The same as in figure \protect\ref{fig2} for the R-region.}
\label{fig5}
\end{figure}

For a fixed location of the brane and at distances from it larger than the
curvature radius, one has $y-y_{0}\gg a$. In terms of the conformal
coordinate this corresponds to $z\gg z_{0}$. If additionally we assume that $%
z\gg L_{i}$, the integral in (\ref{jlRb}) is dominated by the contribution
from the region near the lower limit of the integration and from the mode
with $n_{i}=0$ (under the assumption $|\tilde{\alpha}_{i}|<\pi $), $%
i=p+1,\ldots ,D-1$. By using the asymptotic expression of the Macdonald
function for large arguments, the leading order contribution to the
brane-induced part of the current density is presented as%
\begin{equation}
\langle j^{l}\rangle _{b}\approx \frac{Nez^{D-(p+1)/2}m\tilde{\alpha}_{l}}{%
2^{p+2}\pi ^{(p-1)/2}V_{q}L_{l}a^{D}}\frac{I_{ma+1/2}(z_{0}k_{(q)}^{(0)})}{%
K_{ma+1/2}(z_{0}k_{(q)}^{(0)})}k_{(q)}^{(0)(p-3)/2}e^{-2zk_{(q)}^{(0)}}.
\label{jlbNearH}
\end{equation}%
Note that the large values for $z$ correspond to points near the AdS
horizon. As it has been shown in \cite{Bell17}, in that region the effects
of the gravitational field on the brane-free part of the current density are
small and one has the simple relation $\langle j^{l}\rangle _{0}\approx
(z/a)^{D+1}\langle j^{l}\rangle _{0}^{(\mathrm{M})}$ with the current
density in flat spacetime with toroidal spatial dimensions. This shows that
near the AdS horizon the contribution $\langle j^{l}\rangle _{0}$ dominates
the VEV. In figure \ref{fig5}, the brane-induced current density in the
R-region is depicted as a function of $z/z_{0}$ for separate values of the
ratio $z_{0}/L$ (numbers near the curves) and for fixed values $ma=2$, $%
\tilde{\alpha}=\pi /2$. As seen from the graphs in figure \ref{fig5}, for
fixed values of $z_{0}$ and $z$, near the brane the current density
increases with decreasing $L$ ($z_{0}/L$ increases). At large distances from
the brane the situation is opposite: with decreasing $L$ the current density
decreases. This is also seen from the asymptotic estimate (\ref{jlbNearH})
with $zk_{(q)}^{(0)}=z|\tilde{\alpha}|/L$. From the analysis presented above
it follows that at large distances from the brane the VEV decays as $%
e^{-2(z-z_{0})|\tilde{\alpha}|/L}$ and the sign of $\langle j^{l}\rangle
_{b}/e$ coincides with that for $\tilde{\alpha}$.

Let us consider the VEV of the current density as a function of the location
of the brane. In the limit when the brane is close to the AdS boundary, $%
z_{0}$ is small and in (\ref{jlRb}) we replace the cylindrical functions
with the arguments $uz_{0}$ by their asymptotics for small arguments. The
leading order term is given by%
\begin{eqnarray}
\langle j^{l}\rangle _{b} &\approx &\frac{%
2^{1-2ma}NeA_{p}a^{-D-1}z^{D+2}z_{0}^{2ma+1}}{V_{q}\left( 2ma+1\right)
\Gamma ^{2}\left( ma+1/2\right) }\sum_{\mathbf{n}_{q}}k_{l}k_{(q)}^{p+2ma+2}%
\int_{1}^{\infty }dx\,x^{2ma+2}  \notag \\
&&\times \left( x^{2}-1\right) ^{\frac{p-1}{2}}\left[
K_{ma+1/2}^{2}(xzk_{(q)})-K_{ma-1/2}^{2}(xzk_{(q)})\right] .
\label{jlbRbound}
\end{eqnarray}%
In this limit, for a fixed $z$, the brane-induced contribution decays as $%
z_{0}^{2ma+1}$. For the location of the brane close to the AdS horizon $%
z_{0} $ is large. The limiting case $z_{0}k_{(q)}\gg 1$ has been already
discussed above. The corresponding asymptotic is given by (\ref{jlblim1})
with the replacement $z_{0}-z\rightarrow z-z_{0}$.

The asymptotic behavior of the current density, as a function of the length $%
L_{l}$, is investigated in a way similar to that for the L-region. If $L_{l}$
is much smaller than the lengths of the remaining compact dimensions, the
leading order term coincides with the current density in the model with a
single compact dimension $x^{l}$, when the remaining dimensions are
decompactified. If additionally $L_{l}\ll z$, this term is transformed to
the form that is obtained from (\ref{SmallL1}) with the replacement $%
z_{0}-z\rightarrow z-z_{0}$. With the same replacement in (\ref{SmallL2}),
the asymptotic formula is obtained under the condition $L_{l}\ll z-z_{0}$.
Hence, for small values of $L_{l}$ we have an exponential suppression by the
factor $e^{-2\left( z-z_{0}\right) |\tilde{\alpha}_{l}|/L_{l}}$.

\begin{figure}[tbph]
\begin{center}
\begin{tabular}{cc}
\epsfig{figure=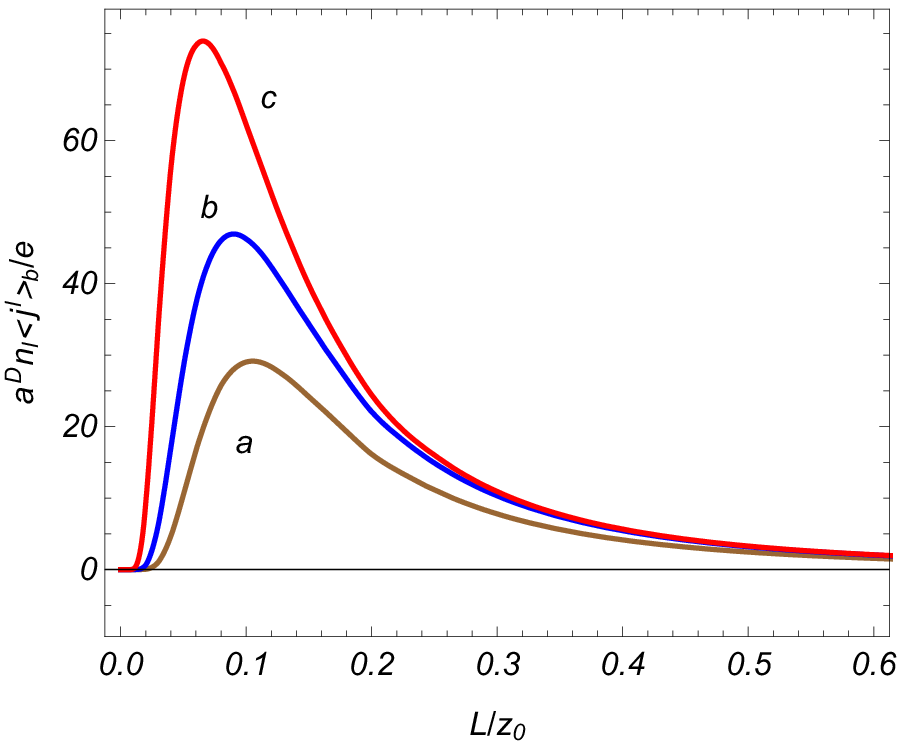,width=7.cm,height=5.5cm} & \quad %
\epsfig{figure=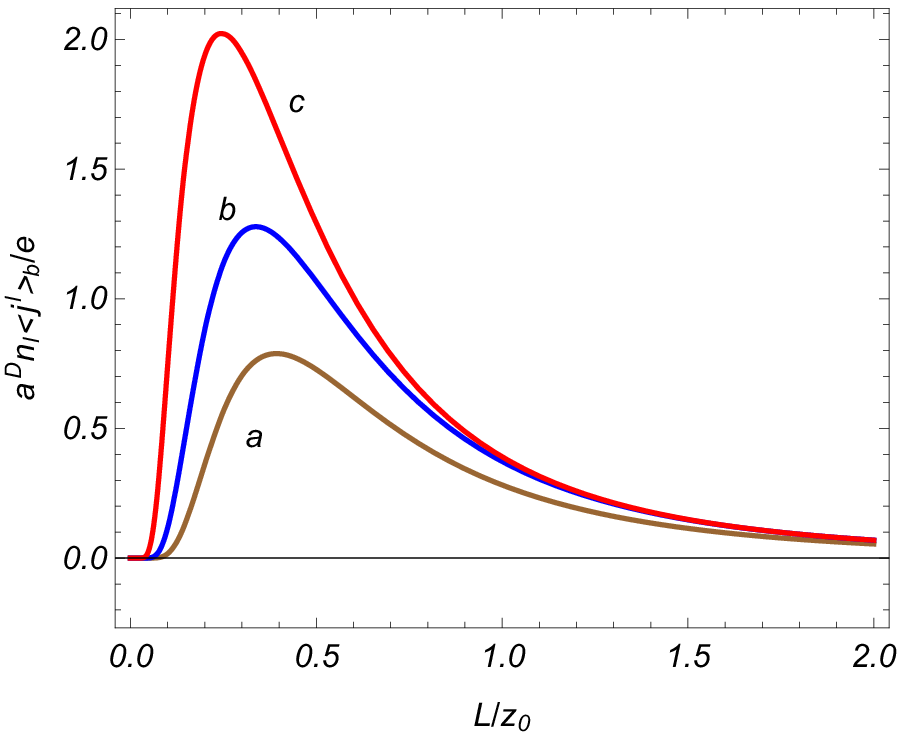,width=7.cm,height=5.5cm}%
\end{tabular}%
\end{center}
\caption{The same as in figure \protect\ref{fig3} in the R-region for $%
z/z_{0}=1.05$ (left panel) and $z/z_{0}=1.2$ (right panel).}
\label{fig6}
\end{figure}

For large values of $L_{l}$ it is convenient to use the representation (\ref%
{jlRalt1}). The argument of the function $g_{p/2+1}(x)$ is large and we use
the corresponding asymptotic expression. Two cases should be considered
separately. For $k_{(q-1)}^{(0)}\neq 0$ (at least one of the phases $\tilde{%
\alpha}_{i}$, $|\tilde{\alpha}_{i}|<\pi $, $i\neq l$, is nonzero), the
integral in (\ref{jlRalt1}) is dominated by the contribution from the region
near the lower limit and by the mode with $r=1$, $k_{(q-1)}=k_{(q-1)}^{(0)}$%
. Using the asymptotics for the Bessel and Neumann functions for small
arguments, the leading order term is expressed as%
\begin{equation}
\langle j^{l}\rangle \approx -\frac{Nea^{-D-1}z^{D+1+2ma}\sin (\tilde{\alpha}%
_{l})k_{(q-1)}^{(0)\left( p/2+ma+1\right) }}{2^{ma+1/2}\left( 2\pi \right)
^{(p+1)/2}\Gamma \left( ma+1/2\right) V_{q}L_{l}^{p/2+ma}}%
e^{-L_{l}k_{(q-1)}^{(0)}},  \label{jllargeLR}
\end{equation}%
with an exponential suppression as a function of $L_{l}$. For $\tilde{\alpha}%
_{i}=0$, $i\neq l$, we have $k_{(q-1)}^{(0)}=0$ and, again, the leading
contribution comes from the mode with $n_{i}=0$, $i\neq l$. Using the
asymptotic expressions for the Bessel and Neumann functions, the integral
with the function $g_{p/2+1}(rL_{l}\lambda )$ in the integrand is expressed
in terms of the gamma function and we get%
\begin{equation}
\langle j^{l}\rangle \approx -\frac{eNa^{-D-1}z^{D+2ma+1}\Gamma \left(
ma+(p+3)/2\right) }{\pi ^{p/2+1}V_{q}L_{l}^{p+2ma+1}\Gamma \left(
ma+1/2\right) }\sum_{r=1}^{\infty }\frac{\sin (r\tilde{\alpha}_{l})}{%
r^{p+2ma+2}},  \label{jllargeLR2}
\end{equation}%
with a power-law decay as a function of $L_{l}$. In particular, this is the
case in models with a single compact dimension. For a massive field this
kind of behavior in a locally AdS bulk is in contrast to that for a locally
flat background geometry, where the decay is exponential, like $e^{-mL_{l}}$%
. Note that the leading terms in both the cases (\ref{jllargeLR}) and (\ref%
{jllargeLR2}) do not depend on the location of the brane and coincide with
the corresponding terms in the brane-free geometry investigated in \cite%
{Bell17}. In figure \ref{fig6} the current density is plotted versus $L/z_{0}
$ for fixed values of $z/z_{0}=1.05$ (left panel) and $z/z_{0}=1.2$ (right
panel) and for the values of the phase $\tilde{\alpha}=2\pi /3$ (a), $\tilde{%
\alpha}=\pi /2$ (b), $\tilde{\alpha}=\pi /3$ (c).

\section{Fermionic currents in $Z_{2}$-symmetric braneworlds}

\label{sec:Brane}

In this section we consider the applications of the results given above in
Randall-Sundrum type braneworlds \cite{Maar10} with a single brane. For the
model described in \cite{Rand99} the background geometry contains two copies
of the R-region, $y<0$ and $y>0$, and the line element is given by (\ref%
{ds2b}) with the warp factor replaced by $e^{-2|y|/a}$. The regions $y<0$
and $y>0$ are related by the $Z_{2}$-symmetry identification $%
y\longleftrightarrow -y$ and $y=0$ is the location of a positive tension
brane. In the original setup the standard model fields are localized on the
brane and there is a single extra dimension $y$ ($D=4$). Most scenarios
motivated from string theories predict the presence of additional bulk
fields and also small extra compact dimensions originating from 10D string
backgrounds. Here we consider a $(D+1) $-dimensional generalization of the
1-brane model with an arbitrary number of toroidally compactified spatial
dimensions. It will be assumed that the brane is located at $y=y_{0}$.

The boundary conditions for the fermionic field $\psi (x)$ on the brane are
dictated by the $Z_{2}$-symmetry of the model. As a consequence of that
symmetry we expect that $\psi (x^{i},y_{0}-y)=M\psi (x^{i},y-y_{0})$ with a $%
N\times N$ matrix $M$. From the invariance of the action under the $Z_{2}$
identification the following conditions are obtained (see also \cite{Flac01b}
for the case $D=4$)%
\begin{equation}
\{\gamma ^{(0)},M\}=0,\;[\gamma ^{(0)}\gamma ^{(b)},M]=0,\;\{\gamma
^{(0)}\gamma ^{(D)},M\}=0,  \label{Mcond}
\end{equation}%
with $b=1,2,\ldots ,D-1$. Now it can be easily checked that these conditions
are satisfied by the choice $M=\zeta \gamma ^{(D)}$. From the condition $%
M^{2}=1$ one gets $\zeta ^{2}=-1$ and, hence, $\zeta =\pm i$. By taking into
account the expression for $\gamma ^{(D)}$ from (\ref{gaml}) we find%
\begin{equation}
M=\mp s\,\mathrm{diag}(1,-1).  \label{M}
\end{equation}%
This matrix is unitary. With the choice (\ref{M}) one gets the same boundary
conditions on the brane for $s=1$ and $s=-1$. For the mode functions (\ref%
{Modesb}) we obtain the boundary condition $Z_{ma+1/2}(\lambda z_{0})=0$ for
the upper sign in (\ref{M}) and the boundary condition $Z_{ma-1/2}(\lambda
z_{0})=0$ for the lower sign.

First let us consider the choice of the upper sign in (\ref{M}). In this
case the boundary condition imposed on the function $Z_{ma+1/2}(\lambda z)$
coincides with that for the bag boundary condition discussed in section \ref%
{sec:Rreg}. Consequently, the expressions for the current density in the $%
Z_{2}$-symmetric braneworld model coincide with those obtained in Section %
\ref{sec:Rreg} with an additional factor 1/2. The appearance of the latter
is related to the fact that in braneworlds the integration over $y$ in the
normalization condition (\ref{Norm}) goes over two copies of the R-region
and, as a consequence, the normalization coefficient for the modes is halved.

For the case with the lower sign in (\ref{M}), from the condition $%
Z_{ma-1/2}(\lambda z_{0})=0$ it follows that the mode functions are given by
\begin{eqnarray}
\psi _{\beta }^{(+)} &=&C_{\beta }^{(+)}z^{\frac{D+1}{2}}e^{i\mathbf{kx}%
-i\omega t}\left(
\begin{array}{c}
\frac{\mathbf{k\chi }\chi _{0}^{\dagger }+i\lambda -\omega }{\omega }%
g_{ma-1/2,ma+s/2}(\lambda z_{0},\lambda z)w^{(\sigma )} \\
i\chi _{0}^{\dagger }\frac{\mathbf{k\chi }\chi _{0}^{\dagger }+i\lambda
+\omega }{\omega }g_{ma-1/2,ma-s/2}(\lambda z_{0},\lambda z)w^{(\sigma )}%
\end{array}%
\right) ,  \notag \\
\psi _{\beta }^{(-)} &=&C_{\beta }^{(-)}z^{\frac{D+1}{2}}e^{i\mathbf{kx}%
+i\omega t}\left(
\begin{array}{c}
i\chi _{0}\frac{\mathbf{k\chi }^{\dagger }\chi _{0}-i\lambda +\omega }{%
\omega }g_{ma-1/2,ma+s/2}(\lambda z_{0},\lambda z)w^{(\sigma )} \\
\frac{\mathbf{k\chi }^{\dagger }\chi _{0}-i\lambda -\omega }{\omega }%
g_{ma-1/2,ma-s/2}(\lambda z_{0},\lambda z)w^{(\sigma )}%
\end{array}%
\right)  \label{Modesbb}
\end{eqnarray}%
with%
\begin{equation}
|C_{\beta }^{(\pm )}|^{2}=\lambda \frac{\left[ J_{ma-1/2}^{2}(\lambda
z_{0})+Y_{ma-1/2}^{2}(\lambda z_{0})\right] ^{-1}}{8\left( 2\pi \right)
^{p}V_{q}a^{D}}.  \label{Cbetb}
\end{equation}%
The reason for the appearance of an additional factor 1/2 in the
normalization coefficients is the same as that for the case of the upper
sign in (\ref{M}). Substituting the mode functions into the formula (\ref%
{jVEV}), by transformations similar to that we have demonstrated in the case
of the bag boundary condition, for the VEV of the current density we find%
\begin{eqnarray}
\langle j^{l}\rangle &=&\frac{1}{2}\langle j^{l}\rangle _{0}-\frac{%
NeA_{p}z^{D+2}}{2V_{q}a^{D+1}}\sum_{\mathbf{n}_{q}}k_{l}\int_{k_{(q)}}^{%
\infty }du\,u\left( u^{2}-k_{(q)}^{2}\right) ^{\frac{p-1}{2}}  \notag \\
&&\times \frac{I_{ma-1/2}(uz_{0})}{K_{ma-1/2}(uz_{0})}\left[
K_{ma+1/2}^{2}(uz)-K_{ma-1/2}^{2}(uz)\right] ,  \label{jlBr}
\end{eqnarray}%
where $\langle j^{l}\rangle _{0}$ is given by (\ref{jl0}).

An alternative representation, obtained by applying the summation formula (%
\ref{AP}), is given by
\begin{eqnarray}
\langle j^{l}\rangle  &=&-\frac{Nea^{-D-1}z^{D+2}}{2\left( 2\pi \right)
^{p/2+1}V_{q}L_{l}^{p}}\sum_{r=1}^{\infty }\frac{\sin (r\tilde{\alpha}_{l})}{%
r^{p+1}}\sum_{\mathbf{n}_{q-1}}\int_{0}^{\infty }d\lambda \lambda   \notag \\
&&\times g_{p/2+1}(rL_{l}\sqrt{\lambda ^{2}+k_{(q-1)}^{2}})\frac{\sum_{j=\pm
1}g_{ma-1/2,ma+j/2}^{2}(\lambda z_{0},\lambda z))}{J_{ma-1/2}^{2}(\lambda
z_{0})+Y_{ma-1/2}^{2}(\lambda z_{0})}\,.  \label{jlBr2}
\end{eqnarray}%
Yet another representation is derived by rotating the integration contour in
a way similar to that described in Section \ref{sec:Rreg}. The corresponding
formula is obtained from (\ref{jlRalt2}) with additional factor 1/2 and by
the replacement%
\begin{equation}
\frac{I_{ma+1/2}(\lambda z_{0})}{K_{ma+1/2}(\lambda z_{0})}\rightarrow -%
\frac{I_{ma-1/2}(\lambda z_{0})}{K_{ma-1/2}(\lambda z_{0})}.  \label{ReplBr}
\end{equation}%
The asymptotic behavior of the current density in various limiting cases is
investigated in a way similar to that in the previous section.

An important difference, compared with the case of the boundary condition
corresponding to the upper sign in (\ref{M}), is the behavior of the
brane-induced VEV in (\ref{jlBr}) for the mass range $ma<1/2$ in the limit
when the location of the brane tends to the AdS boundary, $z_{0}\rightarrow 0
$. In this range of masses we use the relation \cite{Abra72}%
\begin{equation}
\frac{I_{ma-1/2}(uz_{0})}{K_{ma-1/2}(uz_{0})}=\frac{2}{\pi }\cos \left( \pi
ma\right) +\frac{I_{1/2-ma}(uz_{0})}{K_{1/2-ma}(uz_{0})}.  \label{RelIK}
\end{equation}%
The part of the current density corresponding to the last term in (\ref%
{RelIK}) vanishes in the limit $z_{0}\rightarrow 0$ like $z_{0}^{1-2ma}$,
whereas the part with the first term in the right-hand side of (\ref{RelIK})
does not depend on $z_{0}$. Hence, the brane-induced contribution in (\ref%
{jlBr}) (the second term on the right) tends to a finite limiting value:%
\begin{eqnarray}
\lim_{z_{0}\rightarrow 0}\langle j^{l}\rangle _{b} &=&-\frac{NeA_{p}z^{D+2}}{%
\pi V_{q}a^{D+1}}\cos \left( \pi ma\right) \sum_{\mathbf{n}%
_{q}}k_{l}\int_{k_{(q)}}^{\infty }du\,u  \notag \\
&&\times \left( u^{2}-k_{(q)}^{2}\right) ^{\frac{p-1}{2}}\left[
K_{ma+1/2}^{2}(uz)-K_{ma-1/2}^{2}(uz)\right] ,  \label{limBr}
\end{eqnarray}%
with $ma<1/2$. The limiting value of the total current density in this range
of masses is obtained from (\ref{jlBr2}) taking the limit $z_{0}\rightarrow 0
$:%
\begin{eqnarray}
\lim_{z_{0}\rightarrow 0}\langle j^{l}\rangle  &=&-\frac{Nea^{-D-1}z^{D+2}}{%
2\left( 2\pi \right) ^{p/2+1}V_{q}L_{l}^{p}}\sum_{r=1}^{\infty }\frac{\sin (r%
\tilde{\alpha}_{l})}{r^{p+1}}\sum_{\mathbf{n}_{q-1}}\int_{0}^{\infty
}d\lambda \lambda   \notag \\
&&\times g_{p/2+1}(rL_{l}\sqrt{\lambda ^{2}+k_{(q-1)}^{2}})\left[
J_{-1/2-ma}^{2}(\lambda z)+J_{1/2-ma}^{2}(\lambda z)\right] .  \label{limBr2}
\end{eqnarray}%
For a massless field the brane-induced contribution vanishes and in (\ref%
{limBr2}) the brane-free contribution survives only.

We have taken the parameter $\zeta $ in the expression for the matrix $M$
the same for the representations $s=1$ and $s=-1$. We could take $\zeta =\pm
is$ and in this case the matrix $M=\mp \mathrm{diag}(1,-1)$ is the same for
both the representations. However, with this choice, the boundary conditions
on the brane are different for the representations $s=1$ and $s=-1$: $%
Z_{ma+s/2}(\lambda z_{0})=0$ for the upper sign and $Z_{ma-s/2}(\lambda
z_{0})=0$ for the lower one.

\begin{figure}[tbph]
\begin{center}
\begin{tabular}{cc}
\epsfig{figure=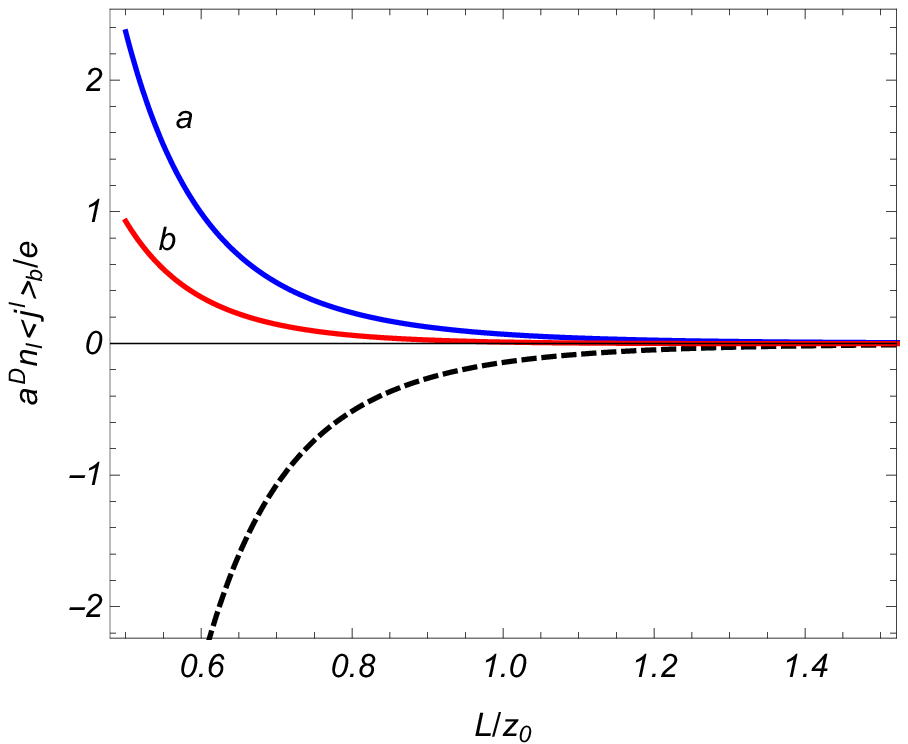,width=7.cm,height=5.5cm} & \quad %
\epsfig{figure=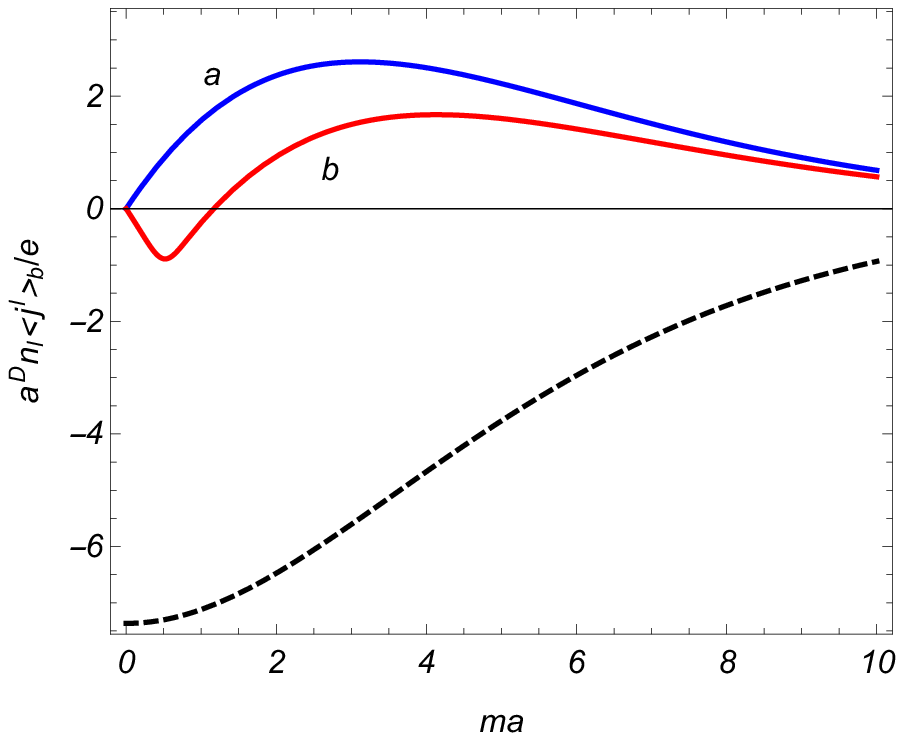,width=7.cm,height=5.5cm}%
\end{tabular}%
\end{center}
\caption{The brane-induced current density on the brane in $Z_{2}$-symmetric
$D=5$ models with a single compact dimension. The curves a and b are for the
upper and lower signs in (\protect\ref{M}). For the left panel we have taken
$\tilde{\protect\alpha}_{l}=\protect\pi /3$ and $ma=2$ and for the right
panel $\tilde{\protect\alpha}_{l}=\protect\pi /3$ and $L/z_{0}=0.5$.}
\label{fig7}
\end{figure}

In Randall-Sundrum type models our universe is realized as a brane and from
the point of view of the interpretation of the results described above by an
observer living on the brane it is of interest to consider the current
density on the brane. For a field with the boundary condition corresponding
to the upper sign in (\ref{M}) the VEV is given by (\ref{jlz0R}) with the
additional factor 1/2. For the boundary condition with the lower sign in (%
\ref{M}) the current density on the brane is directly obtained from (\ref%
{jlBr2}) with $z=z_{0}$:
\begin{equation}
\langle j^{l}\rangle =-\frac{8Nea^{-D-1}z_{0}^{D}}{\left( 2\pi \right)
^{p/2+3}V_{q}L_{l}^{p}}\sum_{r=1}^{\infty }\frac{\sin (r\tilde{\alpha}_{l})}{%
r^{p+1}}\sum_{\mathbf{n}_{q-1}}\int_{0}^{\infty }\frac{du}{u}\frac{%
g_{p/2+1}(rL_{l}\sqrt{u^{2}/z_{0}^{2}+k_{(q-1)}^{2}})}{%
J_{ma-1/2}^{2}(u)+Y_{ma-1/2}^{2}(u)}.  \label{jlonBr}
\end{equation}%
In figure \ref{fig7} we have plotted the brane-induced contributions in the
current density on the brane for the simplest generalization of the
Randall-Sundrum type model with a single extra compact dimension of the
length $L$. In this model $D=5$, $p=3$, $q=1$. The curves a and b correspond
to the upper and lower signs in (\ref{M}). The dashed curve presents the
current density at $z=z_{0}$ in the absence of the brane. The left panel
presents the dependence of the current density on the ratio $L/z_{0}$ for
fixed $\tilde{\alpha}_{l}=\pi /3$ and $ma=2$. On the right panel, the
brane-induced current density is plotted versus the mass (in units of $1/a$)
for the same value of $\tilde{\alpha}_{l}$ and for $L/z_{0}=0.5$. Comparing
the current density on the brane (left panel in figure \ref{fig7}) and
outside the brane (figure \ref{fig6}) as a function of the length $L$, we
see the completely different behavior for small values of $L$. On the brane,
the brane-induced current density increases with decreasing $L$, whereas
outside the brane, for $L<|z-z_{0}|$, it is suppressed by the factor $%
e^{-2|z-z_{0}||\tilde{\alpha}_{l}|/L}$.

\section{Parity and time-reversal symmetric odd-dimensional models}

\label{sec:OddD}

We have considered the fermionic current density for a field realizing the
irreducible representation of the Clifford algebra. In odd-dimensional
spacetimes the mass term $m\bar{\psi}\psi $ in the Lagrangian density is not
invariant under the charge conjugation ($C$) in $D=4n$, under the parity ($P$%
) transformation in $D=4n,4n+2$, and under the time reversal ($T$) in $D=4n+2
$ with $n=0,1,2,\ldots $(of course, $CPT$ invariance holds in all
dimensions, see, for example, Ref. \cite{Shim85}). For even $D$ the matrix $%
\gamma ^{(D)}$ is expressed in terms of the other Dirac matrices as $\gamma
^{(D)}=\pm \gamma $ in $D=4n$ and $\gamma ^{(D)}=\pm i\gamma $ in $D=4n+2$,
with $\gamma =\prod_{i=0}^{D-1}\gamma ^{(i)}$. The upper and lower signs
correspond to the two inequivalent irreducible representations of the
Clifford algebra. One can construct models invariant under the $P$, $C$, and
$T$ transformations combining two $N$-component fermionic fields $\psi _{(s)}
$, $s=\pm 1$, realizing these representations. These fields correspond to
the cases $s=1$ and $s=-1$ in the discussion above. In the gauge $A_{\mu }=0$%
, the Lagrangian density in the model is given by%
\begin{equation}
\mathcal{L}=\sum_{s=\pm 1}\bar{\psi}_{(s)}\left[ i\gamma _{(s)}^{\mu
}(\partial _{\mu }+\Gamma _{\mu }^{(s)})-m\right] \psi _{(s)},  \label{Lag}
\end{equation}%
where $\gamma _{(s)}^{\mu }=(\gamma ^{0},\gamma ^{1},\cdots \gamma
^{D-1},\gamma _{(s)}^{D})$ and $\gamma _{(s)}^{D}=(z/a)\gamma ^{(D)}$ with $%
\gamma ^{(D)}$ from (\ref{gaml}). By suitable transformations of the fields
(see, e.g., Ref. \cite{Shim85}) we can see that (\ref{Lag}) is invariant
under the $C$-, $P$- and $T$-transformations. Now the VEV of the total
current density is the sum of the VEVs coming from the separate fields: $%
\langle J^{\mu }\rangle =\sum_{s=\pm 1}\langle j_{(s)}^{\mu }\rangle $. As
it has been shown above, if the phases in the quasiperiodicity conditions
for the fields $\psi _{(s)}$ are the same then the corresponding current
densities coincide and the total current density is given by the expressions
in Sections \ref{sec:Lreg} and \ref{sec:Rreg} with an additional factor 2.
However, one can have situations where the phases along compact dimensions
are different for separate irreducible representations. An example will be
discussed below.

We can present the Lagrangian density (\ref{Lag}) in terms of a single $2N$%
-component spinor field $\Psi =(\psi _{(+1)},\psi _{(-1)})^{T}$ as%
\begin{equation}
\mathcal{L}=\bar{\Psi}[i\gamma ^{(2N)\mu }(\partial _{\mu }+\Gamma _{\mu
}^{(2N)})-m]\Psi ,  \label{Lag2N}
\end{equation}%
with $2N\times 2N$ Dirac matrices
\begin{equation}
\gamma ^{(2N)\mu }=\left(
\begin{array}{cc}
\gamma _{(+1)}^{\mu } & 0 \\
0 & \gamma _{(-1)}^{\mu }%
\end{array}%
\right)  \label{gam2N}
\end{equation}%
and the related spin connection $\Gamma _{\mu }^{(2N)}$. An equivalent
representation is obtained in terms of the fields $\psi _{(s)}^{\prime }=%
\left[ 1+s+(1-s)\gamma \right] \psi _{(s)}/2$. The combined Lagrangian
density is rewritten as $\mathcal{L}=\sum_{s=\pm 1}\bar{\psi}_{(s)}^{\prime
}[i\gamma ^{\mu }\left( \partial _{\mu }+\Gamma _{\mu }\right) -sm]\psi
_{(s)}^{\prime }$, where now $\gamma ^{\mu }=\gamma _{(+1)}^{\mu }$ and $%
\Gamma _{\mu }$ is the related spin connection. Now the Lagrangian densities
for separate fields are the same except the sign of the mass term. We can
also write the Lagrnagian density in terms of $2N$-component spinor $\Psi
^{\prime }=(\psi _{(+1)}^{\prime },\psi _{(-1)}^{\prime })^{T}$ as $\mathcal{%
L}=\bar{\Psi}^{\prime }[i\gamma ^{\prime (2N)\mu }(\partial _{\mu }+\Gamma
_{\mu }^{\prime (2N)})-m]\Psi ^{\prime }$ with $\gamma ^{\prime (2N)\mu
}=\sigma _{\mathrm{P}3}\otimes \gamma ^{\mu }$, $\Gamma _{\mu }^{\prime
(2N)}=I\otimes \Gamma _{\mu }$.

In view of wide applications of $D=2$ fermionic model in planar condensed
matter systems (so called Dirac materials, including graphene, topological
insulators and Weyl semimetals), here we consider in detail this special
case. The low-energy excitations of the electronic subsystem in these
materials are described by the Dirac equation where the velocity of light is
replaced by a Fermi velocity $v_{F}$. In graphene, the Dirac equation is
written for a 4-component spinor field. The upper and lower two-component
fields of the latter correspond to inequivalent points, $\mathbf{K}_{+}$ and
$\mathbf{K}_{-}$, of the Brillouin zone, whereas their separate components
correspond to the amplitude of the electron wave function on the triangular
sublattices $A$ and $B$ of the graphene honeycomb lattice (for reviews see,
e.g., \cite{Gusy07}). Hence, in the Dirac model describing graphene the
spinor field is presented as $\Psi _{S}=(\psi _{+,AS},\psi _{+,BS},\psi
_{-,AS},\psi _{-,BS})^{T}$, where $S=\pm 1$ corresponds to spin degrees of
freedom. \ The parameter $s=\pm 1$ in our previous discussion corresponds to
the points $\mathbf{K}_{\pm }$. In the special case of $D=2$ bulk geometry
with $\Gamma _{0}^{(4)}=0$, and for the external electromagnetic field with
the potential $A^{\mu }=(\varphi =0,\mathbf{A})$, in standard units the
Lagrangian density reads%
\begin{equation}
\mathcal{L}=\sum_{S=\pm 1}\bar{\Psi}_{S}\left[ i\hbar \gamma ^{(4)0}\partial
_{t}+i\hbar v_{F}\gamma ^{(4)l}\left( \partial _{l}+\Gamma
_{l}^{(4)}-ieA_{l}/\hbar c\right) -\Delta \right] \Psi _{S},  \label{LGraph}
\end{equation}%
where the summation goes over $l=1,2$ and $A_{l}$ is the $l$th component of
the spatial vector $\mathbf{A}$ (for electrons $e=-|e|$). In (\ref{LGraph}),
the energy gap $\Delta $ is included which is related to the Dirac mass by
the formula $\Delta =mv_{F}^{2}$. The corresponding Compton wavelength is
given by $a_{\mathrm{C}}=\hbar v_{F}/\Delta $. Note that, depending on the
gap generation mechanism (see, for instance, \cite{Gusy07}) the mass term
may have more complicated matrix structure. For a given $S$, the Lagrangian
density in (\ref{LGraph}) is an analog of (\ref{Lag2N}) for $D=2$.

Though the topology of the background space for the theory described by (\ref%
{LGraph}) is trivial in a planar graphene sheet, nontrivial topology can be
realized by graphene made structures like carbon nanotubes and nanoloops
with the topologies $R^{1}\times S^{1}$ and $S^{1}\times S^{1}$,
respectively. In both these cases the geometry is flat and the corresponding
ground state currents induced by the threading magnetic flux have been
investigated in \cite{Bell10}. The periodicity conditions on the fields
along compact dimensions of these structures are determined by the chirality
of the tube. In the case of metallic nanotubes and in the absence of
threading magnetic flux, the periodic boundary condition ($\alpha =0$) is
realized for both the fields $\psi _{(\pm 1)}=\left( \psi _{\pm ,AS},\psi
_{\pm ,BS}\right) $. In semiconducting nanotubes for the phases
corresponding to the fields $\psi _{(\pm 1)}$ one has $\alpha =\pm 2\pi /3$
and they have opposite signs for spinors corresponding to the points $%
\mathbf{K}_{+}$ and $\mathbf{K}_{-}$ of the Brillouin zone (see, for
example, \cite{Ando05}). As a result of that, the contributions in the
current density coming from these points cancel each other and, in the
absence of the magnetic flux, the total current density is zero for both
metallic and semiconducting nanotubes.

\begin{figure}[tbph]
\begin{center}
\epsfig{figure=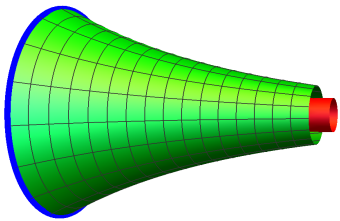,width=5.5cm,height=5.5cm}
\end{center}
\caption{The spatial geometry of deformed $D=2$ tube with an edge, threaded
by a magnetic flux and embedded in three-dimensional Euclidean space. }
\label{fig8}
\end{figure}

The topology in the problem at hand is that realized in carbon nanotubes,
however the tube here is curved (the effects of curvature in graphene made
structures have been discussed in \cite{Kole09,Iori14}). The R-region of
corresponding spatial geometry with an edge, embedded in three-dimensional
Euclidean space, is depicted in figure \ref{fig8}. The magnetic flux
enclosed by the compact dimension is shown separately. A similar type of
spatial geometry has been considered in \cite{Iori14}, though, the geometry
of spacetime differs from that we consider here. Curved graphene tubes with
wormhole geometry were discussed in \cite{Gonz10}.

\begin{figure}[tbph]
\begin{center}
\epsfig{figure=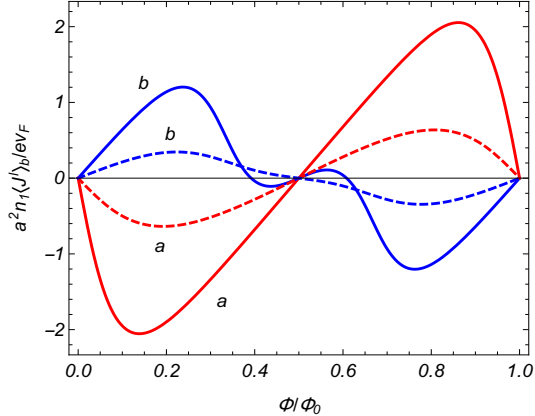,width=7.cm,height=5.5cm}
\end{center}
\caption{The edge-induced current density as a function of the magnetic flux
in $D=2$ metallic (a) and semiconducting (b) tubes for $a/a_{\mathrm{C}}=2$,
$z/z_{0}=1.05$, $z/L=4$ (full curves) and $z/L=2$ (dashed curves).}
\label{fig9}
\end{figure}

In the presence of the magnetic flux, the total current is obtained by
summing the contributions from the fields $\psi _{(+1)}$ and $\psi _{(-1)}$,
given in the previous sections (specified to $D=2$, $q=1$), with the
opposite signs of the phases. As an example, we present here the current
density in the region $z\geq z_{0}$, for a given spin degree of freedom $S$,
obtained from (\ref{jlRalt1}):%
\begin{eqnarray}
\langle J^{1}\rangle  &=&\frac{2ev_{F}z^{4}}{\pi a^{3}z_{0}^{3}}%
\sum_{r=1}^{\infty }\cos (r\alpha )\sin (2\pi r\Phi /\Phi
_{0})\int_{0}^{\infty }du\,u^{2}  \notag \\
&&\times K_{1}(ruL/z_{0})\frac{g_{\nu ,\nu }^{2}(u,uz/z_{0})+g_{\nu ,\nu
-1}^{2}(u,uz/z_{0})}{J_{\nu }^{2}(u)+Y_{\nu }^{2}(u)},  \label{J1D2}
\end{eqnarray}%
where $\alpha $ is the phase for the field $\psi _{(+1)}$, $\nu =a/a_{%
\mathrm{C}}+1/2$ and we have written the mass in terms of the Compton
wavelength. The appearance of $v_{F}$ in the coefficient of (\ref{J1D2}) is
related to fact that the spatial components of the current density for a
given $S$, corresponding to the Lagrangian (\ref{LGraph}), are defined as $%
J^{l}=ev_{F}\bar{\Psi}_{S}\gamma ^{(4)l}\Psi _{S}$. Introducing in (\ref%
{J1D2}) a new integration variable $\lambda =u/z_{0}$ and taking the limit $%
z_{0}\rightarrow 0$, we obtain the current density $\langle J^{1}\rangle _{0}
$ in the edge-free geometry with $0\leq z<\infty $. In this limit, the
integral is expressed as a derivative with respect to $L$ of the integral%
\begin{equation}
\int_{0}^{\infty }d\lambda \,\lambda K_{0}(n\lambda L)\left[ J_{\nu
}^{2}(\lambda z)+J_{\nu -1}^{2}(\lambda z)\right] =\frac{1}{nLz}\left( \frac{%
w-1}{w+1}\right) ^{a/a_{\mathrm{C}}},  \label{Int3}
\end{equation}%
with $w=\sqrt{4z^{2}/\left( nL\right) ^{2}+1}$. It can be checked that the
corresponding expression is reduced to that given in \cite{Bell17}. In
figure \ref{fig9} we display the edge-induced contribution in (\ref{J1D2}), $%
\langle J^{1}\rangle _{b}=\langle J^{1}\rangle -\langle J^{1}\rangle _{0}$,
as a function of the magnetic flux (in units of the flux quantum) for
metallic ($\alpha =0$, curves a) and semiconducting ($\alpha =2\pi /3$,
curves b) quasiperiodicity conditions. The full and dashed curves correspond
to $z/L=4$ and $z/L=2$, respectively. The graphs are plotted for $a/a_{%
\mathrm{C}}=2$ and $z/z_{0}=1.05$.

The fermionic current density induced by a magnetic flux in planar graphene
rings with circular edges has been recently investigated in \cite{Bell16b}.
The fermionic field was confined in the circular ring by the bag boundary
conditions on the edges. In the corresponding problem, in addition to the
azimuthal current, a nonzero VEV of the charge density appears. Note that
persistent currents of a similar physical origin, appearing in normal metal
rings, have been detected in \cite{Bluh09}.

\section{Conclusion}

\label{sec:Conc}

We have investigated the properties of the vacuum state for a charged
fermionic field constrained by two sorts of boundary conditions. The first
one is related to the topology of the background spacetime, being locally
AdS with an arbitrary number of toroidally compactified spatial dimensions.
Along those dimensions the field obeys quasiperiodicity conditions (\ref%
{PerCondb}) with general phases. The toroidal compactification does not
change the local geometry and, because of the high symmetry of the AdS
spacetime, the problem we consider is exactly solvable. The second sort of
boundary condition is imposed by the presence of a brane parallel to the AdS
boundary. We consider the bag boundary condition that is the most popular
condition used for the confinement of fermions in a variety of situations
(an example is the MIT bag model for hadrons). In addition to the background
gravitational field, we also assume the presence of a constant abelian gauge
field. The gauge transformation excluding the latter from the field
equation, leads to the shift of the phases in the quasiperiodicity
conditions for the new field operator. This shift can be formally
interpreted in terms of a magnetic flux enclosed by the compat dimension.
Yet another kind of boundary condition should be imposed on normalizable
irregular modes at the timelike boundary of the AdS spacetime. Here we
consider a special case of allowed boundary conditions corresponding to the
bag boundary condition on a hypersurface close to the AdS boundary with the
subsequent limiting transition to the AdS boundary.

Because of the global nature of the vacuum in quantum field theory, the
expectation values of local physical observables are sensitive to the
boundary conditions on the field. As such a local observable we consider the
fermionic current density. The corresponding VEVs for the charge density and
for the components of the current along uncomapct dimensions vanish. The
brane divides the space into two regions with different properties of the
vacuum and we consider the VEVs for the components of the current along
compact dimensions in these regions separately.

In the region between the brane and AdS boundary the fermionic modes are
presented as (\ref{ModesbL}) and the eigenvalues of the quantum number $%
\lambda $ are quantized by the boundary condition on the brane. In order to
extract from the VEV of the current density the contribution induced by the
brane we have applied to the series over these eigenvalues the summation
formula (\ref{SummAP}). For the component along the $l$th compact dimension
this contribution is given by (\ref{jlLb}). It is an odd periodic function
of the phase $\tilde{\alpha}_{l}$ and an even periodic function of the
remaining phases $\tilde{\alpha}_{i}$, $i\neq l$, with the period equal to $%
2\pi $. For a massless field the result on the locally AdS bulk is
conformally related to the corresponding result on a locally Minkwoski bulk
in the region between two parallel planar boundaries. The latter are
conformal images of the AdS boundary and of the brane. For a massive field,
we have also checked the limiting transition to the corresponding result in
the problem with a single boundary on a locally Minkowski spacetime. On the
AdS boundary, both the brane-free and brane-induced contributions to the
current density tend to zero as $z^{D+2ma+1}$. An important feature that
distinguishes the VEV\ of the current density from the VEV of the
energy-momentum tensor is the finiteness of the former on the brane. For the
investigation of the near-brane asymptotic we have provided an alternative
representation (\ref{jlAlt}). The limiting value of the current density on
the brane is directly obtained from that representation and is given by (\ref%
{jlz0}). The simple relation (\ref{reltot}) takes place between the limiting
value and the integrated current density in the L-region. For $z\gg L_{i}$
and at distances from the brane corresponding to $z_{0}-z\gg L_{i}$ the
brane-induced contribution behaves as (\ref{jlblim2}) and it is mainly
localized near the brane in the region $z_{0}-z\lesssim L_{i}$. For a fixed
observation point, when the brane is close to the AdS horizon, the
brane-induced current density is suppressed by the factor $%
e^{-2k_{(q)}^{(0)}z_{0}}$, with $k_{(q)}^{(0)}$ defined by (\ref{kq0}) for $|%
\tilde{\alpha}_{i}|<\pi $. If the length of the $l$th compact dimension is
much smaller than the lengths of the remaining dimensions, the leading term
in the expansion of the component $\langle j^{l}\rangle _{b}$ coincides with
the current density in the model with a single compact dimension $x^{l}$
when the remaining compact dimensions are decompactified. If, in addition,
one has $L_{l}\ll z_{0}-z$, then the brane-induced current density decays
like $e^{-2\left( z_{0}-z\right) |\tilde{\alpha}_{l}|/L_{l}}$ (see (\ref%
{SmallL2})). This feature is seen in figure \ref{fig3}. In the opposite
limit of large values for $L_{l}$, the brane-induced current density along
the corresponding dimension behaves as (\ref{largeL}) and it is suppressed
by the factor $\exp [-L_{l}\sqrt{\lambda _{1}^{2}/z_{0}^{2}+k_{(q-1)}^{(0)2}}%
]$.

In the region between the brane and the AdS horizon the fermionic mode
functions are given by (\ref{ModesbR}) and the eigenvalues for $\lambda $
are continuous. The brane-induced contribution to the VEV of the current
density in this region is presented in the form (\ref{jlRb}). An alternative
expression for the total current in the R-region is given by (\ref{jlRalt1}%
), with the limiting value on the brane expressed as (\ref{jlz0R}). For a
massless field, the problem is conformally related to the corresponding
problem on a locally Minkowskian spacetime with a single boundary and the
brane-induced contribution vanishes in the R-region. In the case of a
massive field and at large distances from the brane ($z\gg z_{0},L_{i}$,
points near the AdS horizon), that contribution behaves as (\ref{jlbNearH})
and is suppressed by the factor $e^{-2zk_{(q)}^{(0)}}$. In this region the
total current density is dominated by the brane-free part. For a fixed
observation point, when the brane location tends to the AdS boundary, $%
z_{0}\rightarrow 0$, the brane-induced current density tends to zero as $%
z_{0}^{2ma+1}$. If the length $L_{l}$ is much smaller than the lengths of
the remaining compact dimensions and also $L_{l}\ll z-z_{0}$, similar to the
case of the L-region, the brane-induced current decays as $e^{-2\left(
z-z_{0}\right) |\tilde{\alpha}_{l}|/L_{l}}$. In the opposite limit of large
values for $L_{l}$, the asymptotic behavior of the current is essentially
different depending on the phases in the periodicity conditions along the
remaining compact dimensions. If at least one of the phases $\tilde{\alpha}%
_{i}$, $|\tilde{\alpha}_{i}|<\pi $, $i\neq l$, is different from zero, the
leading term in the asymptotic expansion is given by (\ref{jllargeLR}) with
an exponential suppression like $e^{-L_{l}k_{(q-1)}^{(0)}}$. In the case $%
\tilde{\alpha}_{i}=0$, the decay of the current density as a function of $%
L_{l}$ is power law for both massive and massless fields.

In $Z_{2}$-symmetric braneworlds of the Randall-Sundrum type the geometry is
composed by two copies of the R-region related by the $Z_{2}$-symmetry
identification. Depending on the transformation of the field under the $%
Z_{2} $-reflection two types of the boundary conditions on the brane are
obtained. The first one corresponds to the bag boundary condition,
considered in Section \ref{sec:Rreg}. The current density in this case
coincides with that investigated in Section \ref{sec:Rreg} with an
additional factor 1/2, related to the presence of two copies of the
R-region. For the second boundary condition the eigenvalues of the quantum
number $\lambda $ are the zeros of the function $Z_{ma-1/2}(\lambda z_{0})$,
where $z_{0}$ is the location of the brane. The VEV of the current density
in this case is given by (\ref{jlBr}) or, alternatively, by (\ref{jlBr2}).
Now, in the range of the mass $ma<1/2$, the brane-induced current density
does not vanish in the limit when the brane tends to the AdS boundary. The
limiting values for the brane-induced and total currents are given by (\ref%
{limBr}) and (\ref{limBr2}). In the Randall-Sundrum models the observers
reside on the brane and the current density measured by them is presented as
(\ref{jlonBr}). The corresponding dependencies on the length of compact
dimension and on the mass for two types of boundary conditions are
exemplified in figure \ref{fig7}.

In odd-dimensional spacetimes the mass term for a field realizing the
irreducible representation of the Clifford algebra breaks the invariance
with respect two of the $C$-, $P$- and $T$-transformations. Models invariant
under these transformations are constructed by combining the fields
realizing two inequivalent representations of the Clifford algebra. We have
shown that, if the phases in the quasiperiodicity conditions are the same
for these fields, they give the same contribution to the total current
density in this kind of models. From the point of view of applications of
fermionic models in condensed matter physics, an important special case
corresponds to three-dimensional spacetime. An example of physical
realization of those models is graphene. The corresponding low-energy
excitations of the electronic subsystem are described by the Lagrangian
density (\ref{LGraph}). The background topology in this model is nontrivial
for carbon nanotubes and nanoloops (cylindrical and toroidal topologies,
respectively). The phase for the fermionic field along compact dimension of
the nanotube depends on the chirality of the tube. For deformed tubes with $z
$-dependent radius, the resulting current is given by (\ref{J1D2}), where $%
\alpha =0$ and $\alpha =2\pi /3$ for metallic and semiconducting tubes,
respectively. As a consequence of the cancellation of the contributions from
the points $\mathbf{K}_{+}$ and $\mathbf{K}_{-}$ of the Brillouin zone, the
current vanishes in the absence of the magnetic flux $\Phi $ threading the
tube.

\section*{Acknowledgments}

A.A.S. gratefully acknowledges the hospitality of the INFN, Laboratori
Nazionali di Frascati (Frascati, Italy), where a part of this work was done.
V.V.V. acknowledges support through De Sitter cosmology fellowship.

\end{document}